\documentclass[apj]{emulateapj}
\def\approxlt{\lower.2em\hbox{$\buildrel < \over \sim$}}
\def\approxgt{\lower.2em\hbox{$\buildrel > \over \sim$}}

\newcommand{\gsim}{\stackrel{>}{_{\sim}}}
\def\gtrsim{\mathrel{\hbox{\rlap{\hbox{\lower4pt\hbox{$\sim$}}}\hbox{$>$}}}}

\def\lesssim{\mathrel{\hbox{\rlap{\hbox{\lower4pt\hbox{$\sim$}}}\hbox{$<$}}}}
\newcommand{\lsim}{\stackrel{<}{_{\sim}}}
\newcommand{\HyperZ}{{\sc HyperZ}}

\def\la{\mathrel{\hbox{\rlap{\hbox{\lower4pt\hbox{$\sim$}}}\hbox{$<$}}}}
\def\ga{\mathrel{\hbox{\rlap{\hbox{\lower4pt\hbox{$\sim$}}}\hbox{$>$}}}}

\received{????}
\revised{????}
\accepted{????}

\shorttitle{LBGs at $z\sim 2$}
\shortauthors{Haberzettl et al.}

\begin{document}

\title{GALEX selected Lyman Break Galaxies at $z\sim 2$: Comparison with other
  Populations}

\author{L. Haberzettl$^{1}$, G. Williger$^{1,2,3}$, M.D. Lehnert$^{4}$, 
  N. Nesvadba$^{5}$,L. Davies$^{6,7}$}

\affil{$^1$ Department of Physics and Astronomy, University of Louisville, Louisville KY 20492}
\affil{$^2$ Univ. de Nice, Obs. de C\^ote d'Azur, 06108 Nice Cedex 2, France}
\affil{$^3$ IACS, Catholic U. of America, Washington DC 20008}
\affil{$^4$ GEPI, Observatoire de Paris, UMR 8111 du CNRS, 5 Place
Jules Janssen, 92195 Meudon, France}
\affil{$^5$ Institut d'Astrophysique Spatiale, CNRS, Universit\'e Paris-Sud,
Bat. 120-121, 91405 Orsay, France}
\affil{$^6$ Department of Physics, University of Bristol, H H Wills Physics
Laboratory, Tyndall Avenue, Bristol BS8 1TL}
\affil{$^7$ Institute of Cosmology and Gravitation, University of Portsmouth,
  Dennis Sciama Building, Burnaby Road,\\ Portsmouth, PO1 3FX, UK}

\begin{abstract}
We present results of a search for bright Lyman break galaxies at
$1.5\le z\le 2.5$ in the GOODS-S field using a NUV-dropout technique
in combination with color-selection. We derived a sample of 73 LBG
candidates. We compare our selection efficiencies to BM/BX- and BzK
methods (techniques solely based on ground-based data sets), and find
the NUV data to provide greater efficiency for selecting star-forming
galaxies. We estimate LBG candidate ages, masses, star formation rates,
and extinction from fitting PEGASE synthesis evolution models. We find
about 20\% of our LBG candidates are comparable to infrared luminous
LBGs or sub-millimeter galaxies which are thought to be precursors of
massive elliptical galaxies today.  Overall, we can show that although
BM/BX and BzK methods do identify star-forming galaxies at $z\sim 2$,
the sample they provide biases against those star-forming galaxies which
are more massive and contain sizeable red stellar populations.  A true
Lyman break criterion at $z\sim 2$ is therefore more directly comparable
to the populations found at $z\sim 3$, which does contain a red fraction.
\end{abstract}


\section{Introduction}
Since their discovery in the early 1990s \citep[][]{1993AJ....105.2017S},
Lyman break galaxies (LBGs) have been used very successfully
to study galaxy evolution processes in the high redshift
Universe. Using the dropout technique, multi-wavelength surveys
have built up large samples of LBGs at redshifts between $z\sim 3 -
8$, where the Lyman break is accessible with solely ground-based data sets
\citep[e.g.][]{1993AJ....105.2017S,1995AJ....110.2519S,1999ApJ...519....1S,2003ApJ...592..728S,2003ApJ...593..630L,2004ApJ...616L..79B}.
The studies have shown that at $z\gsim 4$, a significant
fraction of the ionizing UV flux is emitted by low luminosity
LBGs \citep[e.g.][]{2003ApJ...593..630L, 2006ApJ...648..299S,
2007ApJ...670..928B,2010Natur.467..940L}.  Analysis of the stellar
populations showed that the population of LBGs at $z\sim5$ consists of
significantly less massive and younger galaxies than comparable samples
at $z\sim3$ \citep[e.g.][]{2007MNRAS.377.1024V,2009ApJ...693..507Y}.
\citeauthor[][]{2007MNRAS.377.1024V} suggested that these LBGs
are experiencing their first significant buildup of stellar mass
\citep[see also][]{2009ApJ...697.1493S}. At $z\sim3$, the LBG population
consists of more massive galaxies that have formed stars over longer
periods \citep[e.g.][]{2001ApJ...562...95S,2008ApJS..175...48R}. A
considerable fraction of the $z\sim 3$ LBGs ($\sim 15\%$) consists
of massive $M_* > 10^{11} M_\odot$ infrared luminous galaxies
\citep[][]{2006ApJ...648...81R,2010MNRAS.401.1521M}; these ILLBGs
are considered one possible class of progenitors of today's massive
elliptical galaxies.

Less is known about the LBGs at $z\lsim 2.5$. At this
redshift, the Lyman break occurs in the UV, which can only be observed
from space. First studies using deep images observed by the 
GALEX satellite obtained samples of LBGs at $z\sim 1$ using FUV dropouts
\citep{2006A&A...450...69B,2007MNRAS.380..986B,2009ApJ...702..506H}.
The majority of LBGs ($\sim$ 75\%) at this redshift are disk-dominated, while
the rest consist of interacting/merging galaxies
\citep[][]{2006A&A...450...69B}. 

At $z\sim2$, a number of groups have worked around UV restrictions by using
selection techniques which solely 
rely on ground based data sets as a proxy method for identifying star-forming
galaxies.  
The BM/BX-method \citep[][]{2004ApJ...607..226A,2004ApJ...604..534S} is based
on {\it U$_n$GR} color selection.  The color cuts were determined from 
a sample of spectroscopically-verified $1.9<z<2.7$ star-forming galaxies, 
which were fitted with dust-reddened SEDs. The resulting best-fits were
redshifted to $z=1.5, 2.0, 2.5$ and used to calculate characteristic colors;
for more details see 
\cite[][]{2004ApJ...607..226A}. 

A second ground-based color method is the BzK-selection
\citep[][]{2004ApJ...617..746D}. The choice of filters covers the flat UV part
of the spectra and flanks the Balmer break at rest-frame 4000~\AA. This method
is sensitive to selecting and distinguishing between
star-forming and passively evolving galaxies at $z\gsim
1.4$. Using the K20 data from the GOODS fields,
\citeauthor[][]{2004ApJ...617..746D} identified 25 star-forming and 7 old
galaxies and reported a high selection efficiency with only 13\% contamination
by interlopers. 

Follow-up studies showed that the optically (BX) selected star-forming galaxies
at $z\sim 2$ have typical stellar masses on the order of
$3\times10^{10}M_{\odot}$ and  typical star formation rates of
$\sim50\,M_{\odot}$~yr$^{-1}$
\citep[e.g.][]{2004ApJ...603L..13R,2005ApJ...626..698S}.    
Space densities are  $\sim 3\times10^{-3}$~Mpc$^{-3}$ with a
correlation length of $r_0 = 4.2\pm0.4 h^{-1}$~Mpc
\citep[][]{2005ApJ...633..748R,2005ApJ...619..697A}, indicating that typical
star-forming galaxies at $z\sim2$ are hosted by $\sim 2\times10^{12}M_{\odot}$
dark matter halos.  

Studies of the optical-NIR selected star-forming galaxies by
\citet[][]{2005ApJ...631L..13D} show that a typical $BzK$ galaxy at $z\sim 2$
is likely to be a ULIRG with $L_{IR} \sim (1-2)\times 10^{12}
L_{\odot}$ and star formation rates SFR $\sim 200 - 300 M_{\odot}
$~yr$^{-1}$. The space density of these galaxies is 
$\sim(1-2)\times 10^{-4}$~Mpc$^{-3}$, which is 2-3 times higher than
at $z\sim0.1$ and 1 \citep[e.g.][]{2003AJ....126.1607S,2005ApJ...632..169L}
and makes them relatively common objects at $z\sim 2$. Their ULIRG nature is 
also supported by morphological studies using HST imaging
\citep[][]{2004ApJ...600L.127D}, indicating that many of the $BzK$-selected
galaxies show hints of merger events driving strong star formation events.

Dynamical studies of star-forming galaxies at $z\sim 2$
\citep[][]{2009ApJ...706.1364F},
together with the results from \citet[][]{2006A&A...450...69B},
indicate that between redshifts of $z\sim 2$ and $z\sim 1$ is when disk
galaxies with significant masses start to contribute to the co-moving
density of galaxies. It is also the phase when the star
formation intensity reaches its peak \citep[][and references
therein]{2004ApJ...615..209H}.   

From these solely ground-based selection techniques and the development of
new instruments like the IFU SINFONI at the VLT and OSIRIS at Keck, we have
learned much about star formation at the critical epoch $z\sim 2$. However,
since the selection processes are different to the techniques used for true
LBGs at $z \gsim 3$, comparisons between the samples are not straightforward. 
A consistent selection method using the Lyman break, independent of intrinsic
galaxy properties, should be used to select consistent samples of star-forming
galaxies at $z\sim 2$. Observations require the space-based facilities like
GALEX or the UV-channels of the newly installed WFC3 on board HST. First
studies include \citet[][]{2010ApJ...720.1708H} and
\citeauthor{2009ApJ...697.1410L}\,  The latter analyzed a very deep
($\sim$140ksec) GALEX NUV image of the Subaru Deep Field
\citep[SDF][]{2004PASJ...56.1011K} and other filters to select true LBGs at
redshifts $1.6<z<2.7$, and compiled a sample of $\sim$8000 LBG candidates down
to $V = 25.3$ with $NUV \approx 27$, including 24 spectroscopically confirmed
at $1.5<z<2.7$. For $z\sim2$ LBGs, they reported a factor $\sim 1.8$
higher summed luminosity density compared to $z\sim 2$ BX and $z\sim 3$ LBGs,
and an increase in the luminosity density by a factor of 3-6 between 
$z\sim 5$ and $z\sim 2$.  

\citeauthor[][]{2010ApJ...720.1708H} examined HST+
WFC3 images (F225W, F275W, and F336W) of a 50 arcmin$^2$
field of the GOODS-S field, as part of an extended
multi-color survey. The group used $U_{225}-, U_{275}-, {\rm and}\,
U_{336}-$dropouts down to $AB=26.5$ combined with color criteria to
select 473 LBG candidates at $z\sim 1.7$, 2.1 and 2.7 respectively. 
The fitted Schechter function parameters agree well with predictions from other
observations with respect to redshift. 

These two studies demonstrate the feasibility of selecting
samples of true LBGs at $z\sim 2$ consistent with higher redshift $z\gsim 3$
LBG samples. Although they are a first step in better understanding this
crucial population of star forming galaxies at an epoch formally known as
the ``redshift desert'', little currently is known about their properties. The
studies either lack data coverage especially in the NIR and mid-IR
\citep[][]{2009ApJ...697.1410L} or field of view available
\citep[][]{2010ApJ...720.1708H}.   

In this paper, we present first results from a dropout+color selection of a
sample of true $z\sim 2$ LBGs based on deep GALEX observations of the
GOODS-S. Our specific purpose is to target a rarer, brighter sample of LBGs
than found in the small, deep survey of \citet[][]{2010ApJ...720.1708H}.
In addition, it complements the wider, shallower study of
\citet[][]{2009ApJ...697.1410L} 
because we use a wider range of filters and target a different field, to
minimize the effects of cosmic variance.
We will show that we can obtain a sample of bright $z\sim 2$ LBGs
similar to the higher redshift population observed at $z \gsim
3$. Here we
report first results of estimates of ages, stellar masses, SFRs, and extinction
for our purely UV-selected $z\sim 2$ LBG sample. Using the deep GALEX
UV data available for the 
CDF-S, we were able to utilize the complete size of the GOODS-S field which
offers a wide range of data from the UV to mid-IR and also tripled
 the  \citeauthor[][]{2010ApJ...720.1708H} sample size. 

The paper is organized as follows: \S2 summarizes the data used, 
\S3 describes our search and defines the selection criteria, and \S4, \S5, and
\S6 present photometric redshifts, SED modeling and the results. Our summary and
conclusions are in \S7. Throughout the paper we use a cosmology with $H_0 =
71$ km s$^{-1}$~Mpc$^{-3}$, $\Omega_M = 0.3, {\rm and}\, \Omega_{\lambda} =
0.7$. All distances quoted are comoving.     

\section{Data}

We base our study on GOODS and supporting surveys, which offer a wide variety
of ground- and space-based imaging and spectroscopic data sets, all centered on
the CDF-S and HDF-N. We concentrated our studies on the CDF-S, since this
field offers the best data coverage. See Table~\ref{data} for a complete
listing.

\begin{figure*}
\includegraphics[width=84mm]{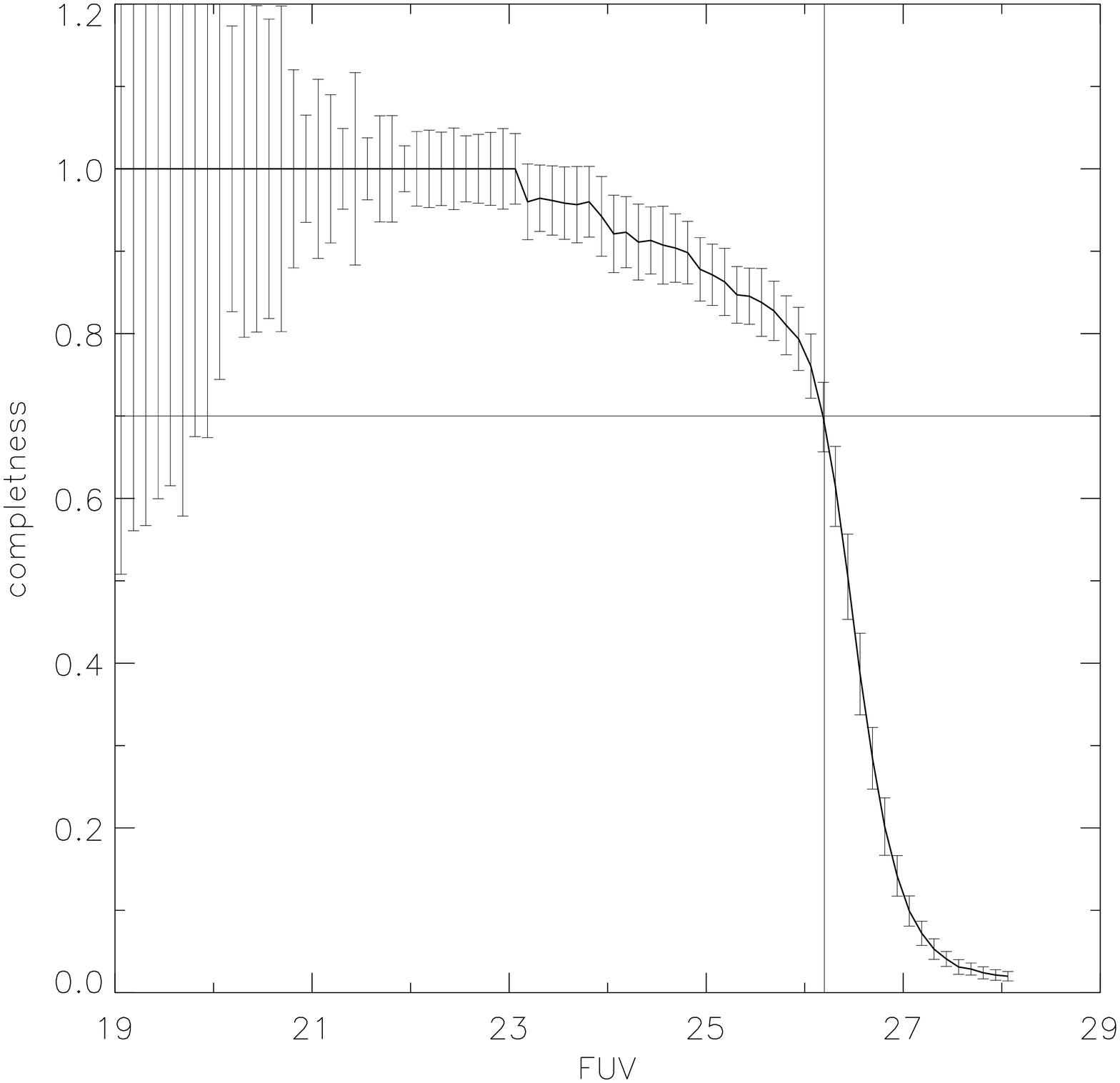}
\includegraphics[width=84mm]{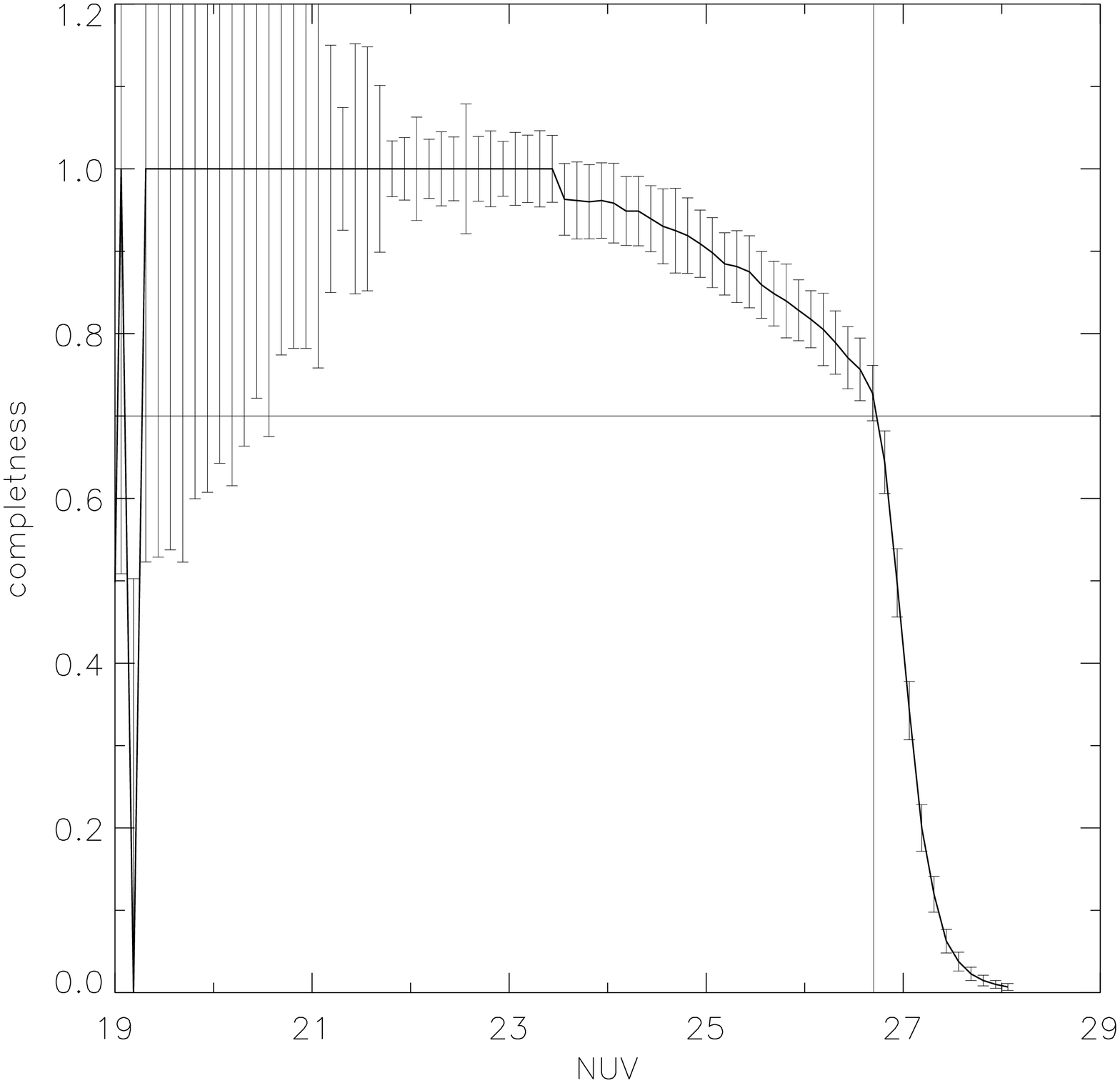}
\caption{\label{nuvhist} Results of the completeness calculations for
  the CDFS GALEX data. {\it Left:} FUV Completeness distribution showing
  a level of 70-75\%  
  at a limit of $\sim$26.2~AB. {\it Right:} NUV completeness
  distribution, indicating a level of  70-75\% for $\sim$26.7~AB.}
\end{figure*}

\subsection{GALEX data}

We used very deep GALEX FUV and NUV imagery, which is part of the GALEX
Deep Imaging Survey (DIS). The data set consists of a 87~ksec FUV and a
89~ksec NUV exposure centered on the CDF-S \citep[][]{2005ApJ...619L...1M}
covering a $\sim$1~deg$^2$ field of view. Without deconvolution, the data
(FUV and NUV) provide a 3~$\sigma$ detection limit of m$_{AB}$=24.5
(MAG\_ISO from SExtractor).  Confusion noise in such deep images, with
the large GALEX PSF ($\sim$5.4$^{\prime\prime}$), contributes to
decreasing the detection sensitivity of the data.  However, since we were
interested in the detection of LBGs at $z\sim$2, where the Lyman break is
shifted in the NUV-band, it is the non-detections and the detection limits
that are important.  Therefore, our results are not seriously affected by
confusion or overlapping sources, and consequently we did not deconvolve the NUV
images.  We adopted an effective detection limit where the completeness
reaches $\sim$ 70-75\% at m$_{AB}$(FUV)$\sim$26.2
and m$_{AB}$(NUV)$\sim$26.7 within a 5$^{\prime\prime}$ aperture
(Figure~\ref{nuvhist}).  We calculated the completeness using
simulated images created with {\it skymaker v3.3.3} and verifying detection probability using {\it
SExtractor}. Figure~\ref{nuvhist} has been
generated using a magnitude bin size of 0.125. The process was repeated
100 times, with 50000 stars per square degree between 18 and 28 mag per
iteration. The errors for the completeness are represented by the standard
deviations for the single bins. The smooth decline in completeness,
starting around $\sim$23 mag, correlates with the number density of
the simulated objects and can be interpreted as result of confusion due
to the large PSF of the GALEX data. The steep decrease 
at faint magnitudes (marked by the big crosses in Figure~\ref{nuvhist})
indicates where the sample starts to become rapidly incomplete.  We chose
a threshold of $\sim 70-75\%$ because it is where the steepest
decline sets in and thus becomes very sensitive
to photometric uncertainties, i.e. the detection or non-detection of sources
becomes sensitive to even small photometric errors.  

To avoid false detections or non-detections (for example due to
objects being merged with bright and/or extended neighbors) we visually
inspected each candidate.  The relatively small number of objects
made this inspection straightforward, and $\sim$10\% of the objects
had multiple identifications which could imply mergers or clumpy extinction
but probably not separate but overlapping sources.

\begin{table*}
\begin{center}
\caption{\label{data} Summary of data used. Unless otherwise noted, m$_{lim}$
  is a 3$\sigma$ detection limit.}
\begin{tabular}{lcccccl}
\hline
Survey&Filter&Telescope&tot. Exp. Time&FWHM&m$_{lim}$&Comment\\
&&&s&arcsec&ABmag&\\
\hline
GALEX DIS&FUV&GALEX&87042&4.5&27.0$^a$&Martin et al. 2005\\
GALEX DIS&NUV&GALEX&89199&5.6&26.5$^a$&Martin et al. 2005\\
GaBoDS & U &ESO/MPG 2.2~m+WFI&78891&1.01&26.3$^b$&Hildebrandt et al. 2005\\
& B & &69431&0.98&27.2$^b$&Hildebrandt et al. 2005\\ 
& V & &104603&0.92&26.8$^b$&Hildebrandt et al. 2005\\ 
& R & &87653&0.79&26.8$^b$&Hildebrandt et al. 2005\\ 
& I & &34575&0.93&25.0$^b$&Hildebrandt et al. 2005\\
GOODS/GEMS&F850LP&HST+ACS&18232&0.03&26.0&Giavalisco et al. 2004,\\
&&&&&&Rix et al. 2004\\ 
GOODS& J &VLT+ISAAC&$\left<15500\right>$&$\left<0.44\right>$&$ <25.5^d$&Giavalisco
et al. 2004\\
& K$_\mathrm{s}$ &data release
v1.0&$\left<20200\right>$&$\left<0.47\right>$&$< 25.2^d$&Giavalisco
et al. 2004\\
& [3.6] &Spitzer+IRAC&82800, 165600$^c$&3.6&$\sim$24.0&Dickinson et
al. in prep.\\
& [4.5] &&82800, 165600$^c$&3.6&$\sim$24.0&Dickinson et
al. in prep.\\
& [5.8] &&82800, 165600$^c$&3.6&$\sim$23.0&Dickinson et
al. in prep.\\
& [8.0] &&82800, 165600$^c$&3.6&$\sim$23.2&Dickinson et
al. in prep.\\
& [24.0] &Spitzer+MIPS&82800&3.6&$\sim$20.5&Dickinson et
al. in prep.\\
& MOS & VLT+VIMOS & $\left<14200\right>$ &&$U_{CTIO} \sim 26$&Popesso et al. 2009 \\
& MOS & VLT+FORS2 &$\left<39000\right>$ & &$z_{850}<26$&Vanzella et al. 2008\\
\hline
\end{tabular}
\end{center}

\hspace{0mm}$^{a}$ 1.5 $\sigma$ detection limit used within SExtractor
search\\ 
\hspace{-115mm}$^{b}$ 5 $\sigma$ detection limit within a 2$^{\prime\prime}$
aperture\\ 
\hspace{-103mm}$^{c}$ total exposure time for the central overlap region\\
\hspace{-103mm}$^{d}$ 10 $\sigma$ detection limit within a
1$^{\prime\prime}$ aperture
\end{table*}

\subsection{Optical and NIR data}
We used deep public images from the GaBods survey for the
object search and color selection of LBG candidates
\citep[][]{2005A&A...441..905H}. The data set offers 0.25 deg$^2$ UBVRI 
pointings centered on the CDF-S, and was observed using the
ESO/MPG 2.2~m telescope with the Wide Field Imager (WFI). 
The images reach 5$\sigma$ detection limits
in UBVRI for a 2$^{\prime\prime}$ aperture of 26.3, 27.2, 26.8, 26.8. 
For more details, see Table~\ref{data}.  

Deep NIR data are available as part of the GOODS survey observed using
ISAAC at the ESO VLT \citep[][]{2004ApJ...600L..93G}. We used data release
v1.0 covering a 10$^\prime \times$16$^\prime$ area observed in the $J$-
and $K_s$-filters and reaching 3$\sigma$ detection limits of $<$25.5
and $<$25.1 respectively. The images yield an average resolution of
$\left< 0.44\right>$~arcsec in $J$  and $\left<0.47\right>$~arcsec
in $K_s$.

Besides the ground based broad-band imaging, we also used high resolution
space-based imaging covered with ACS on the Hubble Space Telescope, in
particular data from GOODS \citep[][]{2004ApJ...600L.103G} as well as GEMS
\citep[][]{2004ApJS..152..163R}, observed in the Sloan z-band. The data
offer a pixel size of 0.03$^{\prime\prime}$ or $\sim$250~pc at $z\sim$2,
and provide us with basic morphological information for
1.5$\le z\le$2.5 LBG candidates.

\subsection{Spitzer data}

Additional photometric information in the mid-IR was provided by
deep Spitzer images observed in the 3.6, 4.5, 5.8, 8.0 and 24$\mu$m
bands. The data are part of the GOODS survey and were observed as a
Spitzer Legacy program in 2004; we used the data version v0.3 (Dickinson
et al., in preparation). The images cover $10^{\prime}\times16^{\prime}$.
Due to the observing strategy for the IRAC bands (3.6 to 8.0 $\mu$m),
which are comprised of 2 exposures taken 6 months apart and rotated
180$^{\circ}$ against each other, the central 1/3 ultra-deep parts of the
final images were exposed twice as long as the outer super-deep regions
(Table~\ref{data}). The MIPS data were only observed during the
epoch 1 observing campaign and correspond to the super-deep IRAC images.
The Spitzer data cover the rest frame NIR for the selected $z\sim 2$
LBGs, and probe the older low mass part of the stellar populations,
thus better constraining the SED fits.

\begin{figure*}
\begin{center}
\includegraphics[width=84mm]{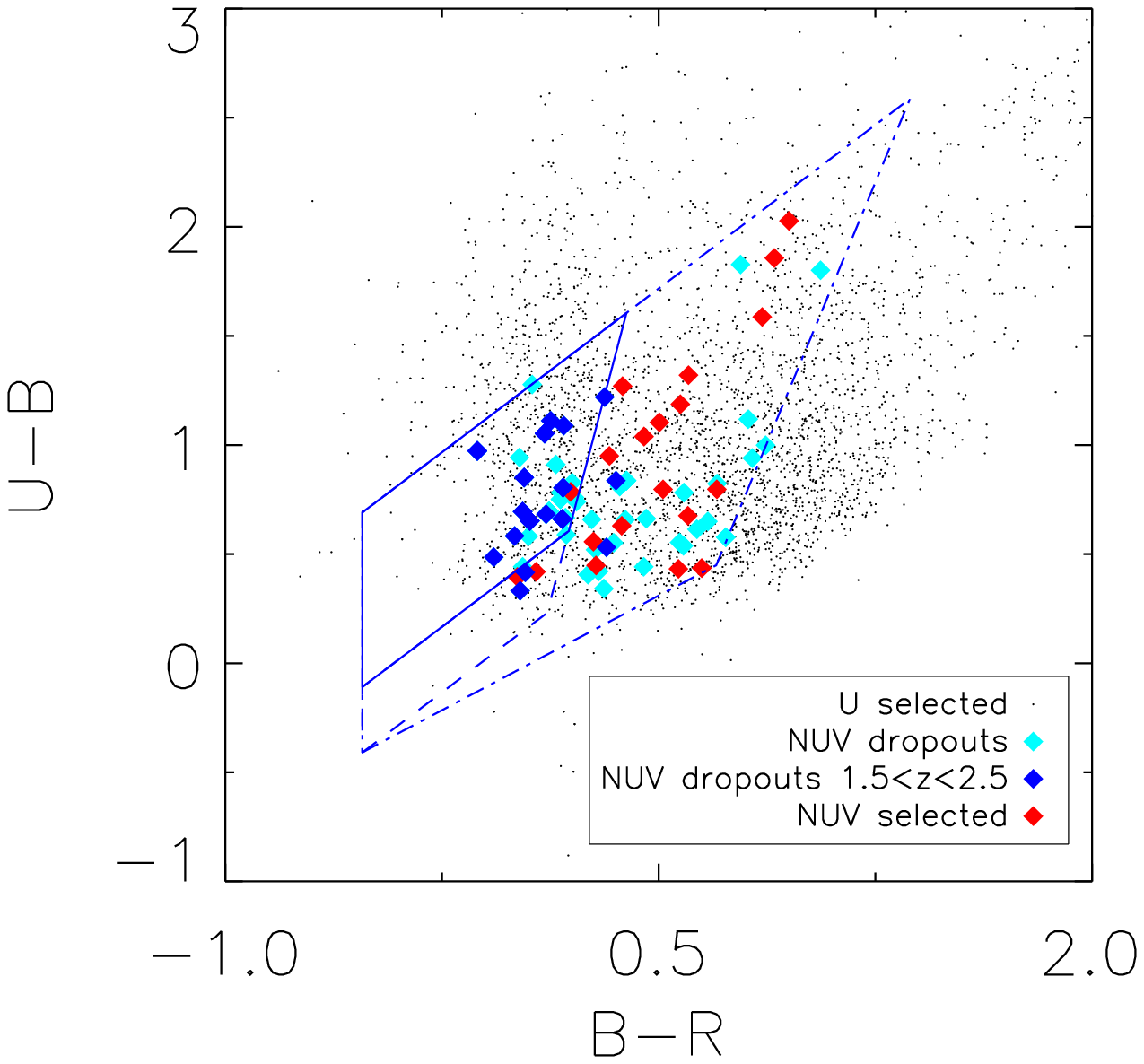}
\includegraphics[width=84mm]{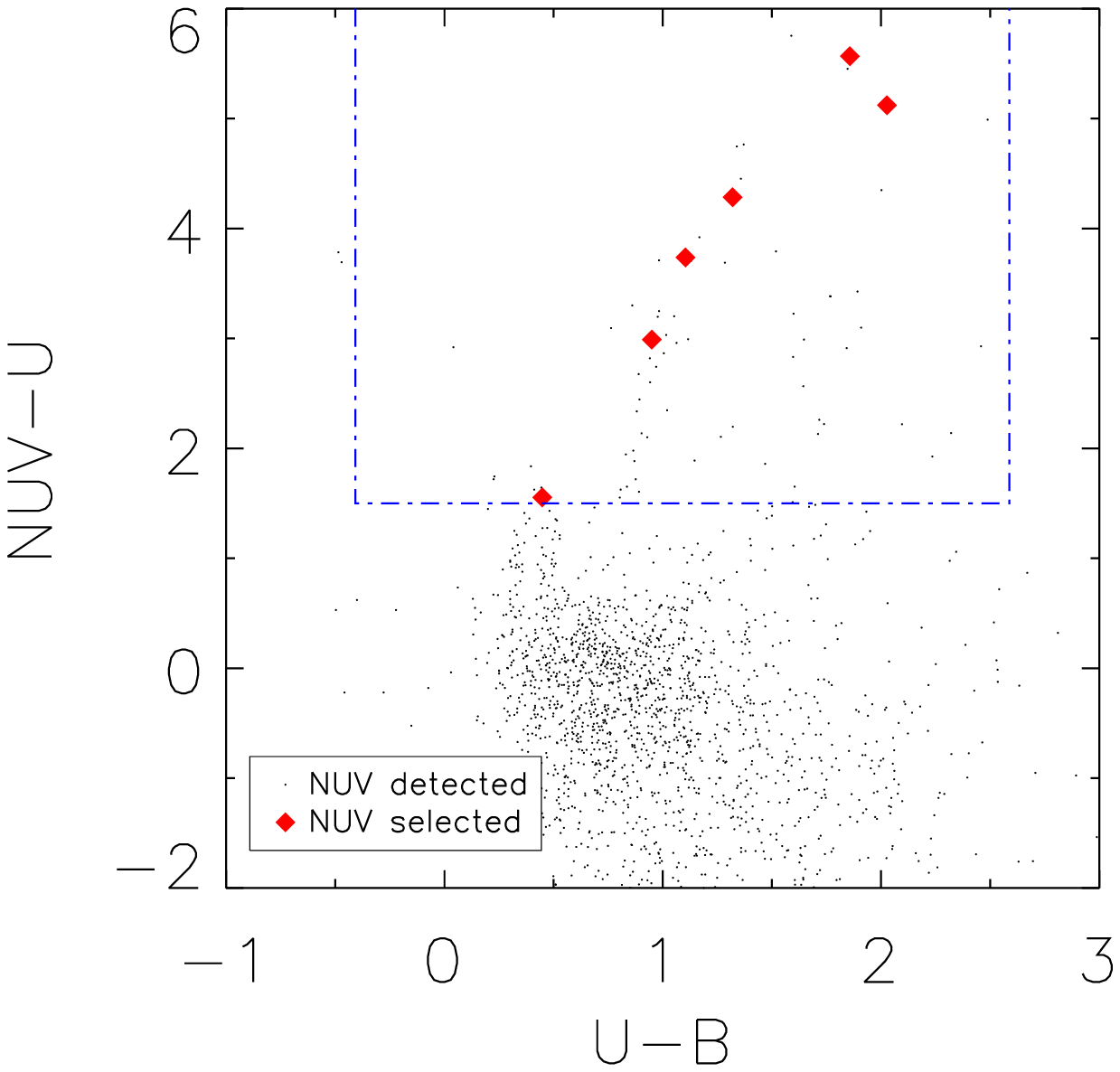}
\includegraphics[width=84mm]{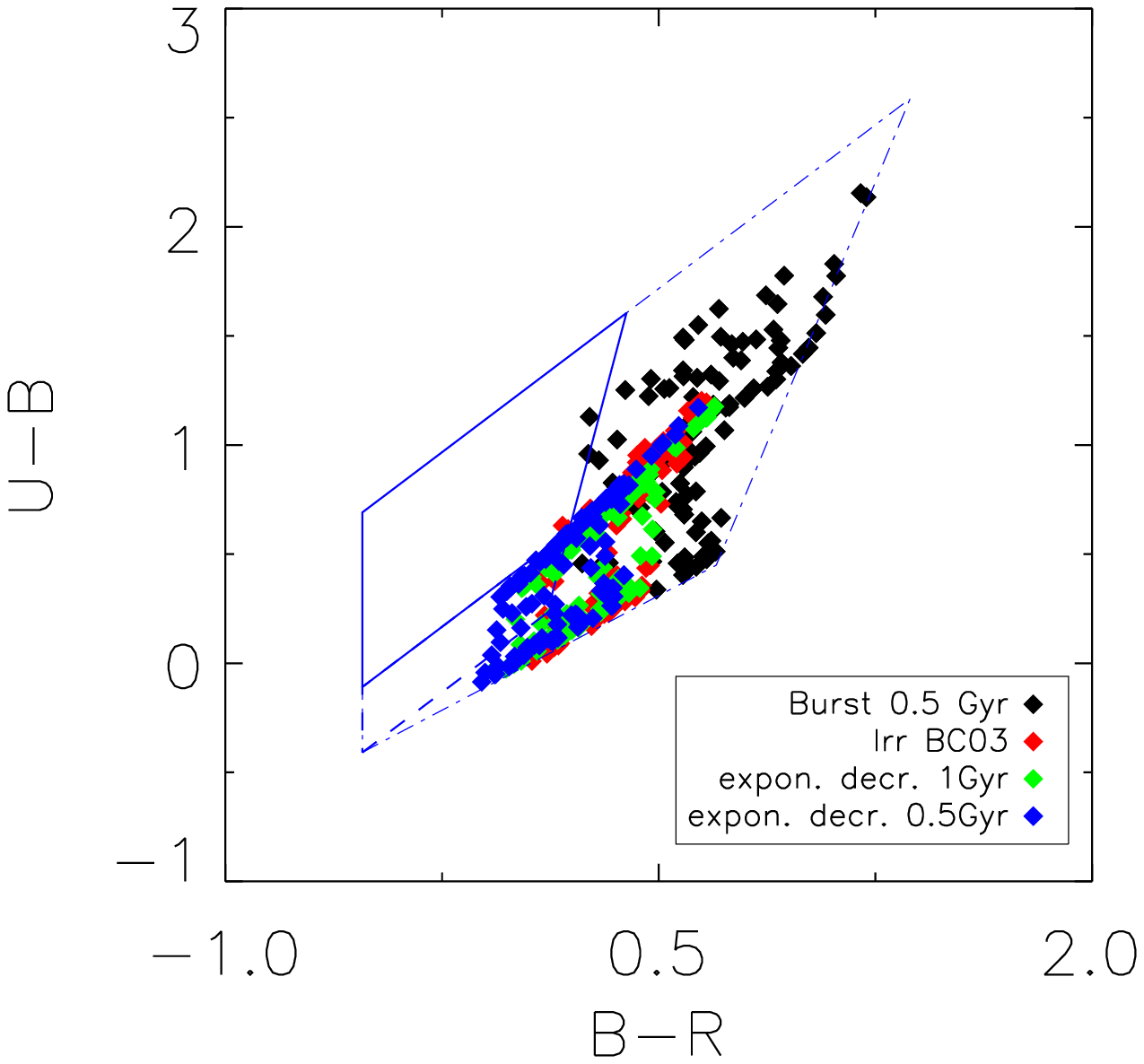}
\includegraphics[width=84mm]{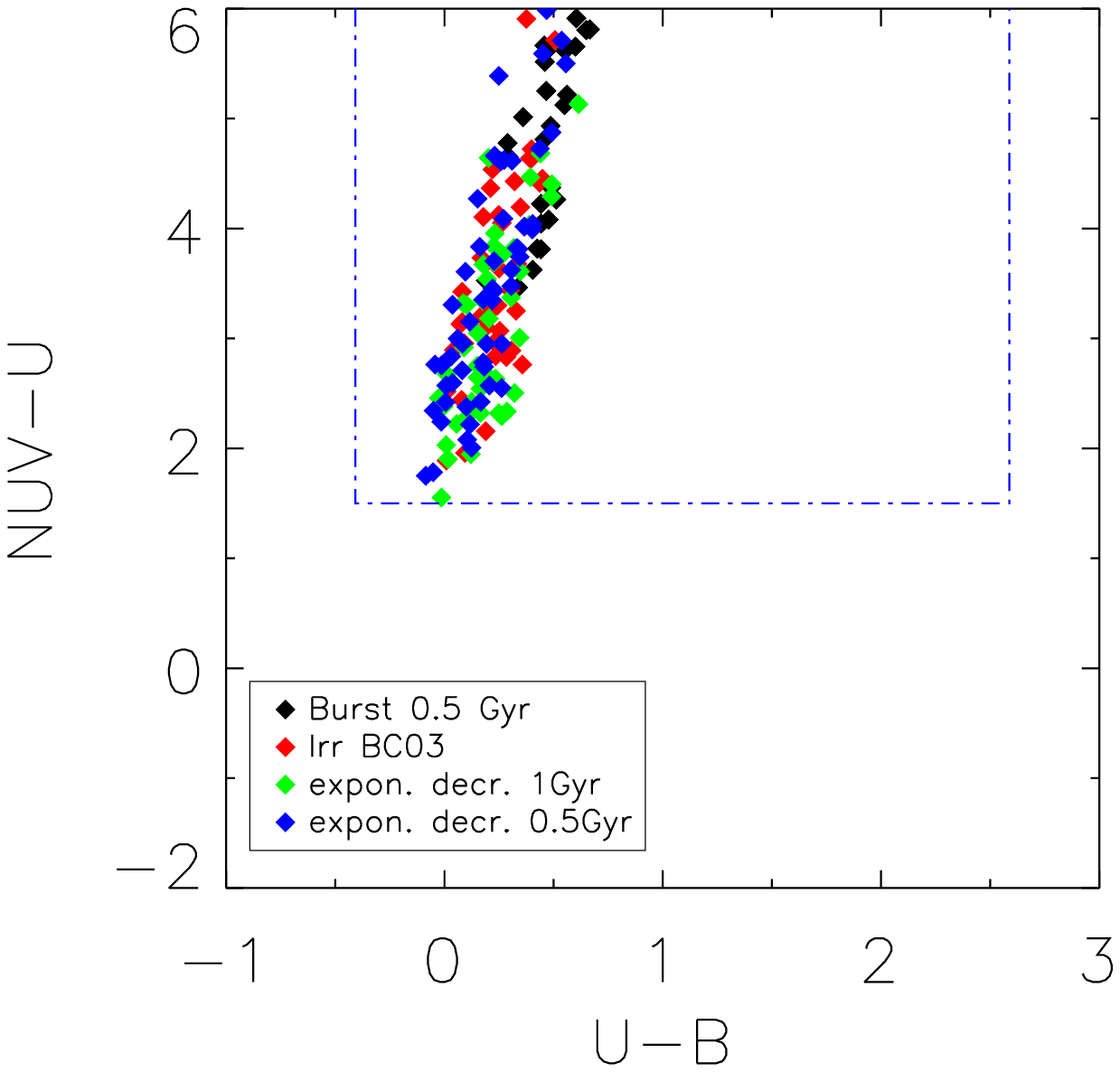}
\caption{\label{colorsel} {\it Lower panels:} Selection regions defined by 
  a series of model colors of  star-forming galaxies in the two color plane, with
  models indicated in the key. 
  {\it Blue dash-dotted quadrilaterals:} Colour selection criteria used for LBG
  candidates at $1.5\le z \le 2.5$ (the largest selection regions, which
  include all other selection regions). {\it Blue solid quadrilaterals, left
  panels:} BX color criteria (optically selected star-forming galaxies for
$2.0<z<2.6$). {\it Blue dashed strip below/blueward of BX region, left panels:} 
  BM color criteria (optically selected star-forming galaxies for
  $1.5<z<2.0$), BM/BX from \citet[][]{2004ApJ...604..534S}.
  We select objects for two groups of LBG candidates. (1) {\it Cyan diamonds,
    upper left panel:} NUV-dropouts with colors complying with the largest
  quadrilateral selection box in the $B-R$ {\it vs.} $U-B$ diagram. 
  (2) {\it Red diamonds, upper two panels:} objects at $1.5\le z \le 2.0$
  detected in the NUV-band and possessing colors complying with the selection
  boxes in both color-color diagrams. {\it Blue diamonds:} A subsample
  of 25 NUV-dropouts with spectroscopically confirmed redshifts 
  at $1.5\le z \le 2.5$. {\it Black dots:} Complete U-band detected sample in
  the GOODS-S field.} 
\end{center}
\end{figure*}

\subsection{Spectroscopic data}

Redshift information is available for parts of our sample from the GOODS-ESO
observations with VLT+VIMOS \citep[][]{2009A&A...494..443P}. 
We used redshifts from the low resolution VIMOS LR-Blue grism data, which
have a spectral resolution of $\sim$28~\AA , with a 5.7~\AA/pixel
dispersion. The spectra cover 3500--6900~\AA\ (1167--2300~\AA\ at $z=2$). 
A second set of redshifts used in our study was also available
via GOODS, and was obtained with the VLT+FORS2 
\citep[][]{2005A&A...434...53V,2006A&A...454..423V,2008A&A...478...83V}.
The observations were taken using the 300I grism without an
order-separating filter, resulting in a 3.2~\AA /pixel dispersion and a 9~\AA\
resolution at 8000~\AA , spanning 6000--10000~\AA. For this study
  we could use a total of 1075 redshifts in the field, covering 0$\le z \le$5. 
\begin{figure*}
\begin{center}
\includegraphics[width=168mm]{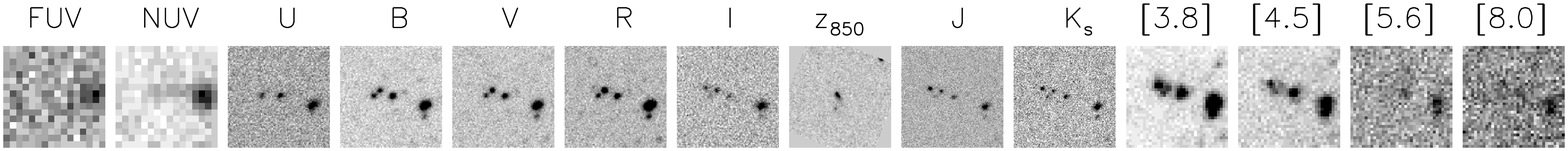}
\includegraphics[width=168mm]{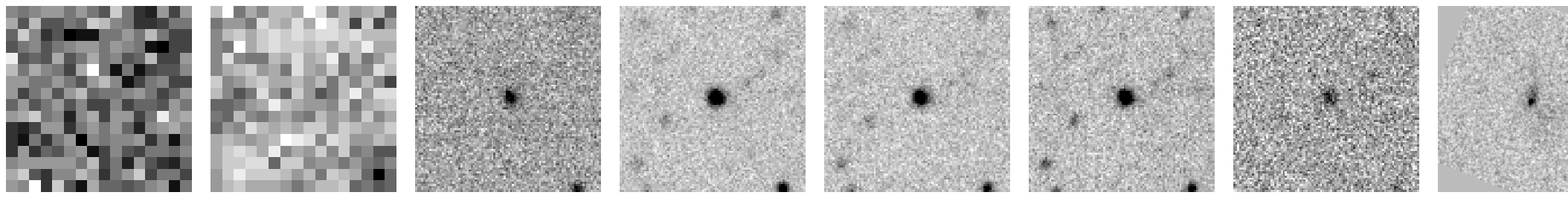}
\caption{\label{example_im} Example images for selected LBG candidates in the
  13 photometric bands (FUV to 8.0$\mu$m) plus the HST-ACS
  z$_{850}$-band. All images are 20$^{\prime\prime}$ in size, except for the
  z$_{850}$-band which is 5$^{\prime\prime}$.}
\end{center}
\end{figure*}

\subsection{Photometry}

In the following, unless stated otherwise, we will solely concentrate
on the analysis of objects found in the GOODS-S, the inner part of the
CDF-S which is covered extensively in all 13 filter bands. Since
the smallest total area coverage ($\sim$140~arcmin$^2$) is offered by
the NIR data observed by ISAAC on the ESO VLT, we used it to determine
the area over which we selected our final sample.


Throughout this paper we  employ both aperture and total magnitudes.
We used the aperture photometry for photometric redshift estimations
and SED fitting. To perform aperture photometry we applied the {\it
aper} task from {\it NASA's IDL Astronomy User's Library}. We
measured the fluxes using apertures with equal radii of 2$^{\prime\prime}$
in the optical and the NIR filter bands, based on double
the FWHM of 1$^{\prime\prime}$ in the U-band which functions as our
detection filter. However, since the GALEX and Spitzer data offer much
lower spatial resolution, we modified the aperture sizes accordingly
(6$^{\prime\prime}$ for GALEX FUV- and NUV-bands and 4$^{\prime\prime}$
for the IRAC bands). We did not determine aperture magnitudes for the
Spitzer $24\mu$m MIPS images.

\begin{table}
\begin{center}
\caption{\label{apercor} Aperture corrections}
\begin{tabular}{lccc}
\hline
Filter&radius&m$_{apercor}$&$\Delta$m$_{apercor}$\\
&$^{\prime\prime}$&ABmag&ABmag\\
\hline
FUV&3.0&0.98&0.78\\
NUV&3.0&0.81&0.48\\
U&1.0&0.35&0.17\\
B&1.0&0.28&0.07\\
V&1.0&0.27&0.07\\
R&1.0&0.23&0.08\\
I&1.0&0.31&0.10\\
J&1.0&0.14&0.11\\
K$_S$&1.0&0.14&0.17\\
$[3.6]$&2.0&0.34&0.10\\
$[4.5]$&2.0&0.21&0.08\\
$[5.8]$&2.0&0.17&0.06\\
$[8.0]$&2.0&0.31&0.17\\
\hline
\end{tabular}
\end{center}
\end{table}


%
%

To account for aperture losses, we derived aperture corrections
in every filter-band using a set of about 100 stars found in the GOODS-S
field. The aperture corrections and their uncertainties range between
$0.98\pm 0.78$ for the GALEX FUV and $0.14\pm0.17$ for the ISAAC K$_S$
(Table~\ref{apercor}), and  were calculated from the growth
curves using the mean aperture magnitudes and standard deviations
of the stellar sample. The total errors for the aperture magnitudes were
determined by adding the uncertainties of the magnitude zero-points and
aperture corrections to the errors determined by {\it aper} in quadrature.
For the Spitzer IRAC-bands we used the following magnitude zero-point
errors: $\Delta m_{3.6} = 0.05$~mag, $\Delta m_{4.5} = 0.05$~mag, $\Delta 
m_{5.8} = 0.04$~mag, $\Delta m_{8.0} = 0.06$~mag, which were
calculated from the flux uncertainties of the flux zero-points given in
the Spitzer manual.

We adopted total magnitudes from the SExtractor $MAG\_ISO$-magnitudes.
These total magnitudes were used for the color-selection of the
LBG sample (see \S~\ref{colsel}) and the IR color-magnitude diagrams
in \S~\ref{illbg}.

\section{Search and color selection}
\label{colsel}

To select LBG candidates in the redshift range 1.5$\le z\le$2.5, we first
detected possible candidates performing SExtractor searches in
every single filter band. Specifically, we did not extract sources
from one image based on the detection of an object in another image.
Rather, we detected objects in every
filter band independently, and then measured total magnitudes for them. 
The resulting object catalogs were then
cross-correlated against each other, creating a master catalog including
all imaging results. For the correlation, we applied matching radii
according to the spatial resolution of the images used (Table~\ref{data},
column 5). As a final step, we matched the imaging master catalog with
the two spectroscopic catalogs from GOODS (VIMOS and FORS spectroscopy).

%
%

To identify LBG candidates at $1.5 \le z\le 2.5$, we employed
two different approaches. We limited the size of the
field searched to the area of the sky covered by the ISAAC NIR
observations. Since this data set covers the smallest area, this
step guaranteed to restrict our sample only to candidates observed
in all 13 filter-bands. Over this field we selected all U-band sources
brighter than 25.3 (AB magnitudes) with no detection in the FUV-
and NUV-bands (NUV-dropouts i.e. $NUV \lsim 26.7$ in a
5$^{\prime\prime}$ aperture), and compared their locations in a $U-B$
vs. $B-R$ color-color diagram to those of model galaxies at $1.5 \le
z\le 2.5$ (Figure~\ref{colorsel}). The U-band selection limit
allowed us to detect objects with a Lyman break (NUV-U) of at least
1.4 magnitudes.

We calculated model galaxy colors using template SEDs from
\citet[][PEGASE]{1997A&A...326..950F} and \citet[][]{2003MNRAS.344.1000B}, and
employed the photo-z algorithm \HyperZ\ 
\citep[][]{2000A&A...363..476B}  to derive the model color tracks
(Table~\ref{modelpar})  for a variety of 
star formation laws, ages and dust properties (Figure~\ref{colorsel} bottom
row). The template SEDs were extinction-adjusted 
via \HyperZ\ using the reddening law of
\citet[][]{2000ApJ...533..682C} and $A_V$ ranging between 0 and 1.2. We
furthermore accounted for absorption by the Ly$\alpha$ forest on the SEDs using the method
from \citet[][]{1995ApJ...441...18M} within \HyperZ .    
\begin{table}
\caption{\label{modelpar}Template SEDs used for deriving color selection
  criteria (Figure~\ref{colorsel}).}
\begin{tabular}{l l c}
\hline
model&star formation law&age\\ 
&&[Gyr]\\
\hline
Irr Bruzual \& Charlot 2003&constant&2-3.7\\
PEGASE&burst&0.5\\
&exponential decr.&0.5\\
&exponential decr.&1.0\\
\hline
\end{tabular}
\end{table}

The location of the model galaxies in the $U-B~vs.~B-R$ diagram defined the
color selection criteria (largest blue dash-dotted quadrilaterals
in Figure~\ref{colorsel}) for
$z\sim 1.5-2.5$ galaxies
which, with the NUV Lyman break, we used to select our LBG candidates
(Figure~\ref{colorsel} top row). The color cuts are:
\begin{eqnarray}
\nonumber
&NUV &> 26.70\\
\nonumber
&B-R &\ge -0.53\\
\nonumber
&U-B &\ge 0.70 (B-R) - 0.04\\
\nonumber
&B-R &\le 0.31 (U-B) + 0.56\\
&U-B &\le (B-R) + 1.22
\end{eqnarray}
We applied a second set of selection criteria for LBGs at the
lower redshift end ($1.5\lsim z\lsim 2.0$), for which the redshifted
Lyman break falls within the NUV band ($2280{\rm ~\AA} \lsim \lambda \lsim 2736
{\rm ~\AA}$).  Those color cuts are (Figure~\ref{colorsel}, left panels):
\begin{eqnarray}
\nonumber
&FUV &> 26.2\\
\nonumber
&NUV-U& \ge 1.50\\
\nonumber
&U-B& \ge -0.41\\
&U-B& \le 2.59
\end{eqnarray}
The NUV-selected LBG candidates had to meet both sets of color
selection criteria, and be detected in the NUV but not in the FUV. The $U-B$
limits in the $NUV-U$ vs. $U-B$ diagram were the same as in the
$U-B$ vs. $B-R$ diagram, while the $NUV-U~\ge~1.5$ limit represents
the blue end of the model SED tracks.  

Finally, we cleaned our selected sample for stellar contamination 
by comparing it with a stellar catalog
\citep[][]{2002A&A...392..741G}, which was created using a
magnitude-limited, SED fitting approach. This resulted in a sample of
$\sim$400 stellar candidates in the central 0.1~deg$^2$ of the
CDF-S. About 10\% of the NUV-dropout and 47\% of the NUV-selected
galaxies were rejected as stars.  
As a further check,
we visually inspected the NUV images of the LBG candidates and
  rejected objects which were false negatives (non-detections in the
  NUV), e.g. objects merged with bright neighbors.
  In total, the cleaning for stellar contamination and the 
  visual inspection for false NUV-dropouts  reduced our sample
  size from 171 to 127 candidates.  

For our subsequent analysis, we concentrate on the bright population of LBGs in our
target redshift range. Therefore, we only included objects which
were observed in all filter bands between U and 4.5$\mu$m. That
resulted in a sample with uniform sampling across a wide wavelength
range for SED and
photometric redshift determinations.
This final sample contains 53 NUV-dropout LBG
candidates (see Figure~\ref{example_im} for 2 example images in the available 
filter bands) and 20 NUV-selected LBG candidates down to $U\le25.3$.
\label{red_phot}
\begin{figure*}
\includegraphics[width=58mm]{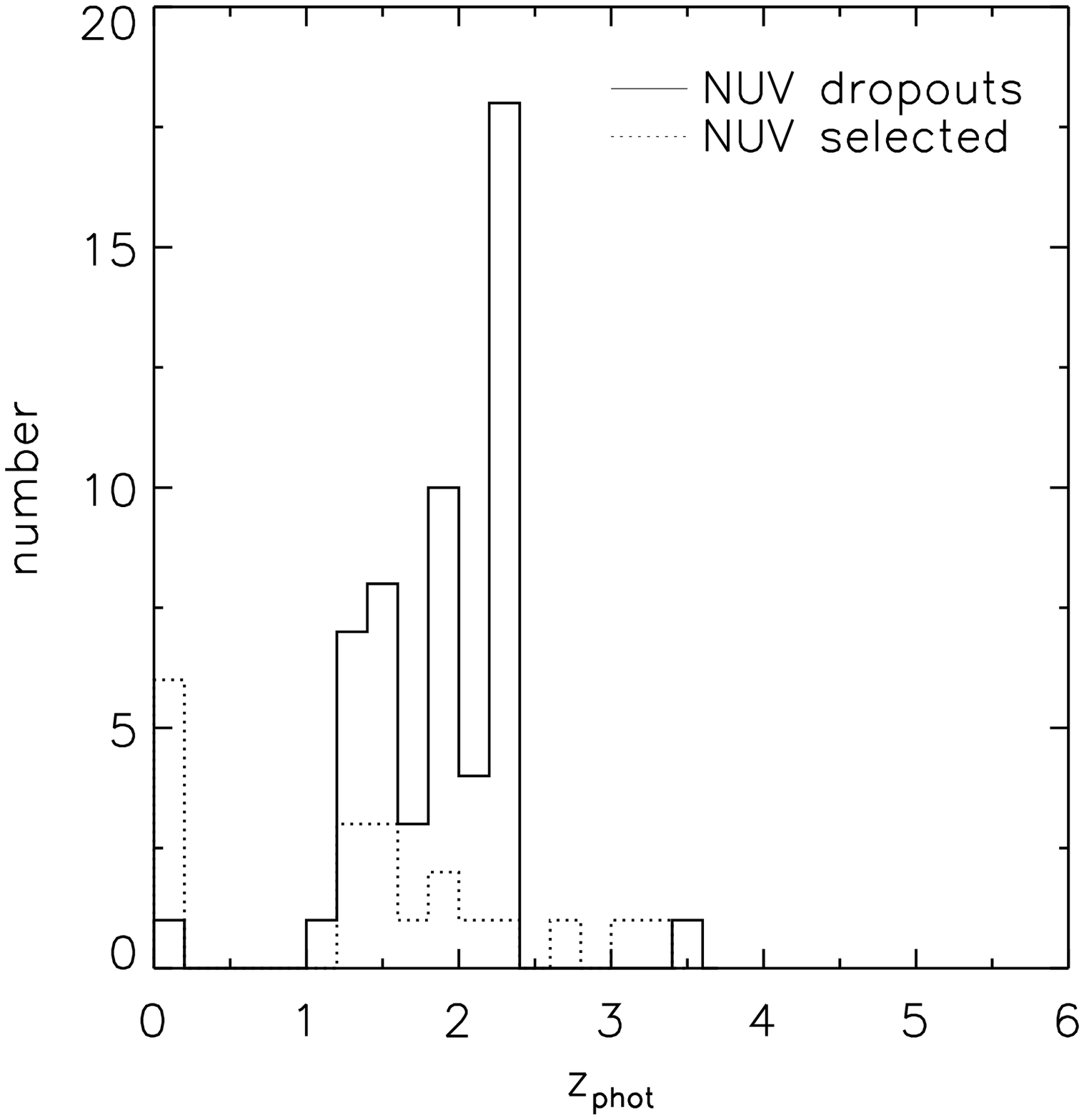}
\includegraphics[width=58mm]{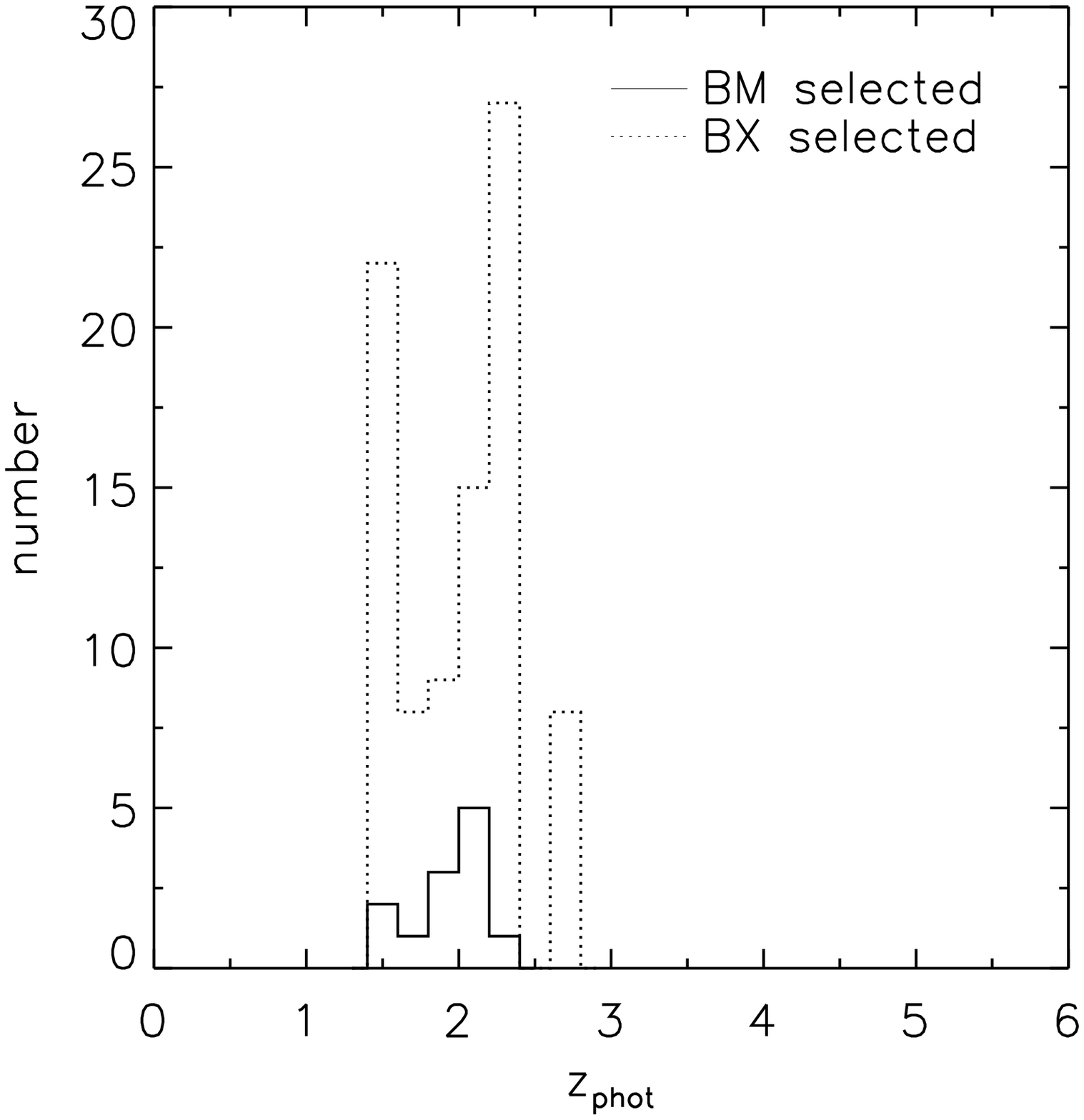}
\includegraphics[width=58mm]{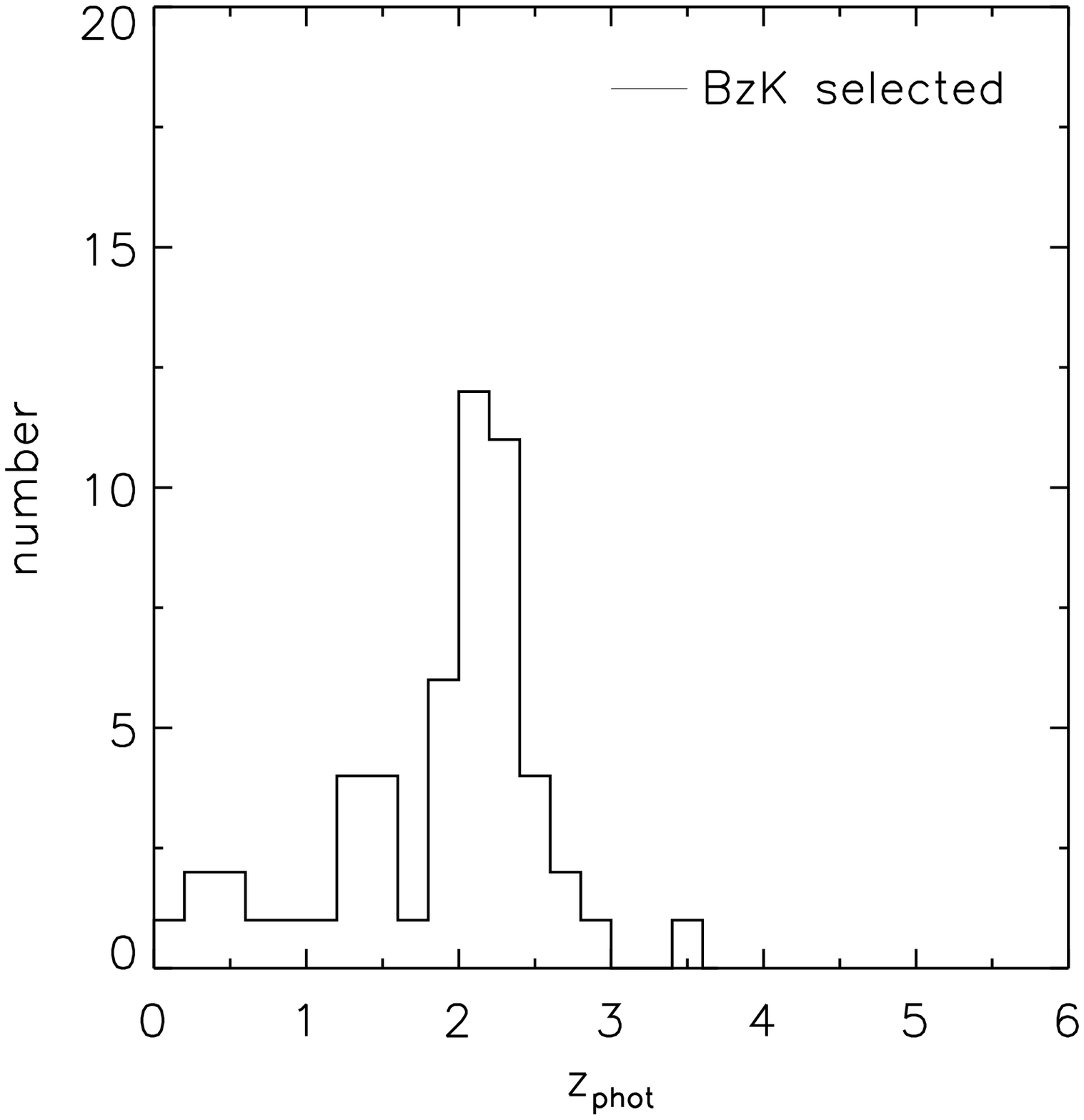}
\caption{\label{photozall}The panels show the photometric redshift
  distributions for star-forming galaxies at $z\sim 2$, selected using the NUV
  dropout (left), BM/BX (middle) and BzK (right) methods.
  The low redshift peak (at $z\sim 0$) in the left panel 
  is mainly due to residual stellar contamination.} 
\end{figure*}


\section{Photometric Redshifts}

Using our 13-band photometry from UV to IR (GALEX FUV to Spitzer 8$\mu$m, but
excluding the ACS z-band), we estimated photometric redshifts with
\HyperZ~v1.3 \citep[][]{2000A&A...363..476B}. The redshift determination was
done by cross-correlating a set of template spectra to the colors of
our sample galaxies. In the current version of \HyperZ , we used
a set of five template spectra from \citet[][]{2003MNRAS.344.1000B},
consisting of an elliptical, Sc, Sd, Irr, and starburst galaxy. 
In cases where the objects were not detected in a filter band, except for the
GALEX FUV- and NUV-bands, we omitted the non-detection for the redshift
determination.  To account for the Lyman break, the flux levels in the GALEX
FUV and NUV band non-detections were set to zero during the template
correlation. \HyperZ\ can account for the effect of the Ly$\alpha$/Ly$\beta$
forest, therefore we applied the corresponding Ly$\alpha$ forest opacity 
estimates of \citet[][]{1995ApJ...441...18M}. 
We accounted for internal reddening of the LBG candidates by
employing  the reddening law of \citet[][]{2000ApJ...533..682C},
with $A_V$ set to range between 0 and 1.2 mag in steps of 0.2.  
The resulting redshift distributions for galaxy samples selected  with
the BM/BX, 
BzK and our dropout technique are shown in Figure~\ref{photozall}.  
The peak at low redshift $z\sim 0$ in the dropout and NUV selected sample
(left panel) is mainly due to residual stellar contamination in the NUV
selected sample. Five stars were identified when checking the HST images
by eye. One of the low redshift galaxies, selected with the NUV selection 
method, clearly shows flux in the FUV filter-band. The flux is weak and 
merged with emission from a nearby bright star, leading to a false
non-detection. Both the star and the galaxy were removed from the sample and not
considered for further analysis.
The second low redshift galaxy, selected with the NUV dropout 
method, does not show obvious emission in the UV filter-bands. We therefore
keep this object in our sample at this stage. Our subsequent analysis will use redshift limits to
restrict the
sample further, minimizing the effects on the results by
foreground or background objects. 

While our NUV dropout selection and the BM/BX method allowed for
a relatively clean selection of $z\sim 2$ star forming galaxies, the
contamination with low redshift interlopers is much higher for the
$BzK$ selection method.  

\begin{figure*}
\includegraphics[width=170mm]{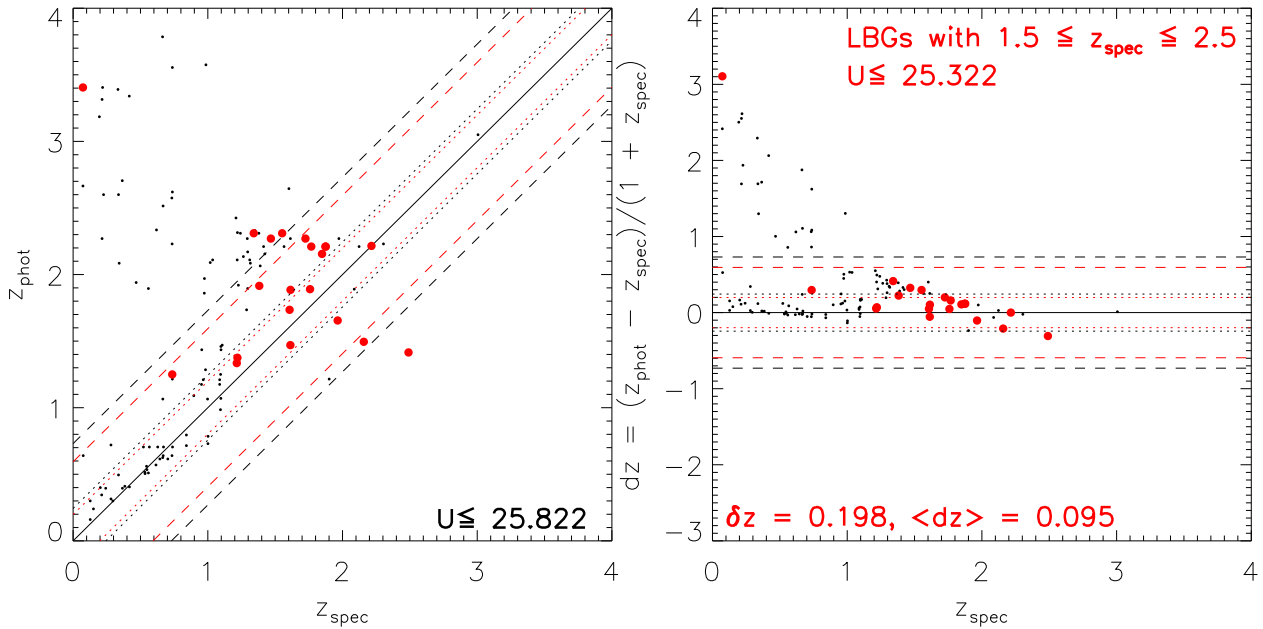}
\caption{\label{redshift} {\it Left:}
  Comparison between photometric redshifts $z_{phot}$ calculated using
  template SED fitting algorithm \HyperZ\ and spectroscopic
  redshifts from VLT+VIMOS and FORS observations. {\it Solid line:}
  $z_{phot} = z_{spec}$ in order to guide the eye. {\it Dotted, dashed lines:}
  1$\sigma$  and 3$\sigma$ deviations. 
  {\it Black dotted, dashed lines:} $1\sigma$, $3\sigma$ limits for the 
  complete galaxy sample.  {\it Red dotted, dashed lines:} $1\sigma$,
  $3\sigma$ limits for the 
  NUV-dropout selected LBG sample. {\it Red circles:} Our LBG sample.
  {\it Right:} The normalized photometric redshift accuracies $dz_{offset} =
  (z_{photo} - z_{spec})/(1+z_{spec})$. Line types and symbols are the same as
  in the left panel.  Our LBG sample has an accuracy of $\delta z_{rms} =
  0.198$ and a mean offset of $\left<dz_{offset}\right> = 0.095$.} 
\end{figure*}

We calculated photometric redshift uncertainties $dz_{offset} = (z_{phot} -
z_{spec})/(1 + z_{spec})$ (right panel Figure~\ref{redshift}). Overall, the rms
is $\delta z_{rms} = 0.24$, with an averaged offset of $\langle
dz_{offset}\rangle = 0.13$. The fraction of catastrophic outliers ($\delta
z_{rms} > 3\sigma$) is 14\%. The accuracy of the photometric redshift
estimation increases with the use of our LBG selection method:
for our LBG candidate sample, the rms is $\delta z_{rms} = 0.198$, with an
average of $\langle dz_{offset}\rangle = 0.095$, indicating no
significant systemic offsets. The fraction of catastrophic
outliers decreased to $\sim$5\% (1/20). We therefore conclude that the
photometric redshift determination enables us to estimate unbiased photometric
redshifts for our LBG sample.  

\section{SED Modeling}

In order to estimate intrinsic properties of the $1.5<z<2.5$ LBG sample,
we fitted our 13-band aperture photometry to a library of spectral
energy distributions (SEDs) created by the synthesis evolution
model PEGASE \citep[][]{1997A&A...326..950F}.  We chose the PEGASE
models mainly because of the possibility to model SEDs using consistent
chemical evolution which constrains the metallicity as a function of time.
We think this is a more realistic approach than using stellar populations
with a fixed metallicity \citep[e.g.][]{2003MNRAS.344.1000B}. Also, due to
an otherwise similar modeling approach (similar stellar library, similar
treatment of the TP-AGB phase, etc.) our results are more comparable to
other studies of $z\gsim 2$ LBGs based on \citeauthor{2003MNRAS.344.1000B}
models \citep[e.g.][]{2001ApJ...562...95S,2005ApJ...626..698S}.

Our library consisted of several thousand PEGASE spectra including
star formation histories with constant and exponentially decreasing star
formation rates. The decay times $\tau$ for the exponentially decreasing
SFRs range from $\tau = 10$~Myr (burst-like scenarios) to $\tau =
5000$~Myr (close to constant star formation).

We modeled the SEDs using a \citet[][]{1993MNRAS.262..545K} initial mass
function (IMF), including consistent chemical evolution and following
a closed box approach excluding gas infall and outflow. 
Although that it is likely that high-redshift star-forming galaxies have
outflows \citep[e.g.][]{2010ApJ...717..289S}, we decided not to consider
outflows in the PEGASE SEDs. Again, this allows for a better comparison with
previously published results based on the \citeauthor{2003MNRAS.344.1000B}
models.

The model fits and the conversion into physical parameters (mass,
SFR, $E(B-V)$) were done using an IDL based chi-square minimization
routine. Our routine applies the \citet[][]{2000ApJ...533..682C}
extinction law to the PEGASE models for dust correction. We
chose the \citeauthor[][]{2000ApJ...533..682C} law since it
has been used by many other studies of high redshift galaxies
\citep[e.g.][]{2001ApJ...562...95S,2005ApJ...626..698S,
2009ApJ...693..507Y}.  We applied extinction corrections
to the PEGASE model SEDs ranging between A$_V$=0-5~mag in steps of
0.1~mag. The extinction-corrected PEGASE models were then redshifted
and fitted following the chi-square minimization approach described
by \citet[][]{2005AJ....130.1337G}. The normalization of the measured
SEDs was derived including only the observed frame optical ($UBVR$)
filter-bands.

When possible, we employed spectroscopic redshifts, and photometric redshifts
otherwise.  

From the model fits we estimated ages, stellar masses, SFRs and dust
extinction $E(B-V)$ for LBG candidates. Our results only allow
for best-fitting SEDs which were younger than the age of the Universe.
We estimated uncertainties for the calculated parameters as the standard
deviation of the expectation value of each parameter \citep[see][for
more details]{2009A&A...507.1793N}. We restrict the final sample of
LBG candidates to galaxies with photometric and spectroscopic
redshifts $1.0\le z\le 2.5$, which includes 61 objects.

%
%
     
     
The SED fitting resulted in a median of the reduced $\chi^2$ values
for the complete sample $ \chi^2 = 1.83\pm 1.70$. Acceptable fits
with a probability of 
0.1\% (1\%) have $\chi^2 \leq 3.25 (2.5)$ and make up about 70\%  (56\%) of
the sample.  
Larger reduced $\chi^2$ values appear to result from two main sources:\\   
1) small photometric errors, especially in the NIR and mid-IR filter bands,
indicating that we are underestimating the photometric uncertainties
in these bands.\\  
2) Uncertainties in the photometric redshift estimation. We
find that reduced $\chi^2$-values for objects with spectroscopic redshifts
($\sim 34\%$) are smaller than for objects with only photometric redshifts
available. The median reduced $\chi^2$ for the objects with spectroscopic
redshifts is $\chi^2 = 1.18\pm 1.53$, with about 83\% having
$\chi^2 \leq 3.25$, while the median reduced $\chi^2$ for the objects with
only photometric redshifts is $\chi^2 = 2.63\pm 1.70$. This
indicates that uncertainties in the photometric redshift estimation resulted in
higher uncertainties in the SED fits. Since even small redshift uncertainties
can have large implications for the fit result when including the NUV filter
(especially for the lower redshift LBGs $z\sim 1.5$), we decided to exclude
the NUV and FUV photometry from the SED fits.

The majority of best-fitting models are those with either constant
SFRs (11\%) or long decay times $\tau = 1500-5000$~Myr (61\%) which
are in practice nearly constant. Another 20\% of the fitted galaxies are best
described by exponentially decreasing SFRs with moderate decay times
$\tau =150-1000$~Myr. Only 8\% of the galaxies are best represented
by exponentially decreasing SFRs with short decay times $\tau = 10 - 100
$~Myr i.e. starbursts.  All except
one of the fitted galaxies are best described by dust-reddened
SED models. About 84\% of the sample is best fitted with an extinction
correction of  $A_V > 0.5$. and 52\% with $A_V > 1$.

Stellar masses and SFRs are relatively well constrained with median
uncertainties of about -0.4~dex. We did not find an obvious correlation
between stellar masses and SFRs {\it vs.} their $\chi^2$-values and their
errors. However, the highest masses ($>10^{10}$M$_\odot$) seem to
be best constrained, showing the smallest uncertainties. In contrast,
the uncertainties for the ages are large, with a median value
of 0.9~dex. Older galaxies in the sample ($>500$~Myr) are
found to have relatively well-constraint ages with a median error of
$\sim$0.4~dex. Unlike the stellar mass and the SFRs, the ages show an increase
in uncertainties with increasing $\chi^2$.  Therefore, most of the uncertainties in
the best-fitting models lie in the age estimates.

\begin{figure*}
\includegraphics[width=180mm]{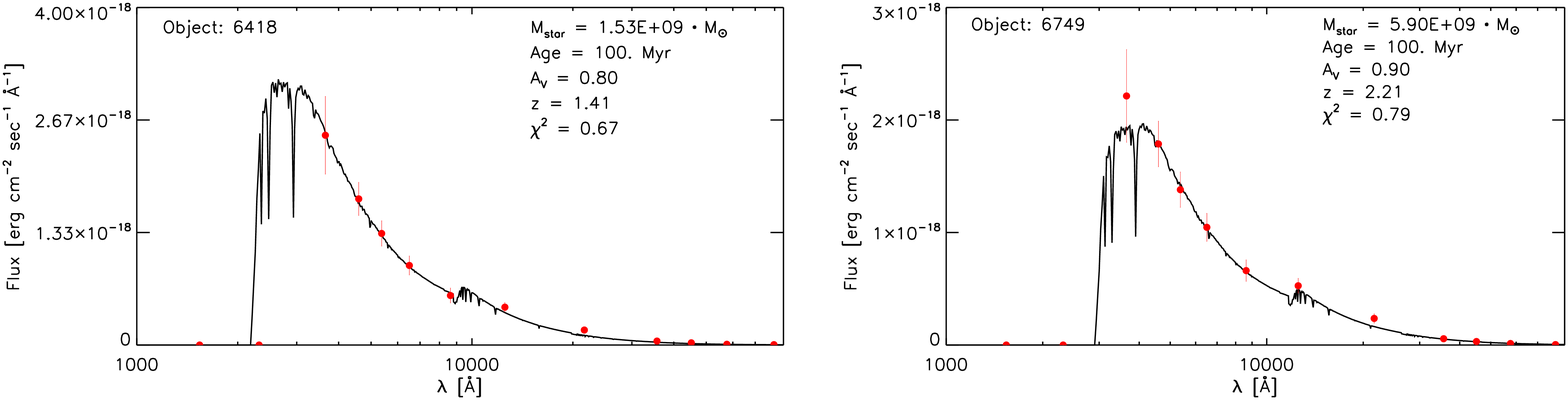}
\caption{\label{SEDexamp} Example SED-fits of a PEGASE library of
  synthesis evolution models  fitted using our $\chi^2$-minimization
    algorithm. They represent fits} with spectroscopic {(\it
    left)} and photometric {(\it right}) redshifts. Error-bars for
  the $R,J,K_S,$ and IRAC-bands are smaller than the symbol sizes.
\end{figure*}

Examples of successful SED fits are shown in Figure~\ref{SEDexamp}. 
The right panel shows an SED for which only photometric
redshifts were available, while the left panel shows an example fit for
which the spectroscopic redshifts could be used.   
\label{sedfits}


\section{Results}
\subsection{Dropout selection}
From the GOODS-S spectroscopic data
\citep[][]{2008A&A...478...83V,2009A&A...494..443P}, we were able to obtain
spectroscopic 
redshifts for 25 (20 dropout and 5 NUV-selected) of our 67 total 
LBG candidates. About 80\% of the dropout selected (16/20) and
only 20\%  (1/5) of the NUV selected LBG candidates were found to have
redshifts in our target range of $1.5\lsim z\lsim 2.5$. However, an additional
$\sim$15\% (3/20) of the NUV-dropouts and 60\% (3/5) of the 
NUV-selected objects have
$1.0<z<1.5$. Including this poorly explored redshift range, inaccessible
from the ground for true dropout identifications, increased 
the selection efficiency from $\sim$80\% to 95\%
for the dropout selected and 20\% to 80\% for the NUV selected LBG
candidates. 

{\it Comparison with BM/BX.}
In the last few years, several selection methods like the BM/BX-
\citep[][]{2004ApJ...604..534S,2004ApJ...607..226A}
and BzK-methods \citep[][]{2004ApJ...617..746D,2004ApJ...600L.127D} have been
employed, which solely rely on ground-based photometry to create
samples of star-forming galaxies at $z\sim 2$. We compared the
selection efficiency of these methods
here to our dropout selection which directly probe
their Lyman break spectral regions, and thus are more directly analogous
to the LBG samples at $z>3$. To avoid effects due to selecting fainter 
objects especially with the BM/BX method, we applied the same brightness
limits ($U \leq 25.32$~AB) as for the dropout selection. 
Figure~\ref{colorsel} shows our Lyman break+color
criteria (large dot-dashed quadrilaterals, left panels)
compared to the BM/BX-method (smaller solid/dashed blue quadrilaterals 
respectively).  The overlap is relatively small, and consists 
mainly of model galaxies with relatively complex star formation histories
(e.g. exponentially decreasing SFRs).  The allowed LBG parameter space 
increases when including more evolved and/or dust-reddened cases
-- because the dropout method does not depend on (optical) colors
redward of the rest-frame Lyman limit.
In summary, only 2/3 of the NUV-dropout selected sample of  
LBG
candidates (34 of 53) are detected with the BM/BX criteria, and only about 60\%
of the BM/BX selected galaxies (26 of 45) would have been selected by our
color plus NUV-dropout criteria. 

{\it Comparison with BzK.}
Only 9 LBG candidates ($\sim$20\% of the NUV-dropout selected LBG
candidates) satisfy the BzK ``$B-z\,{\rm vs.} \,z-K_s$" color requirement.
Only $\sim$25\% of the BzK-selected galaxies (11 of 48) were found to be
NUV-dropouts and about $\sim$80\% of the BzK NUV-dropouts (9 of 11)
were located within our color-selection region in the $U-B\,
{\rm vs.} \,B-R$ diagram (Fig.~\ref{colorsel}). \\

\begin{figure*}
\begin{center}
\includegraphics[width=168mm]{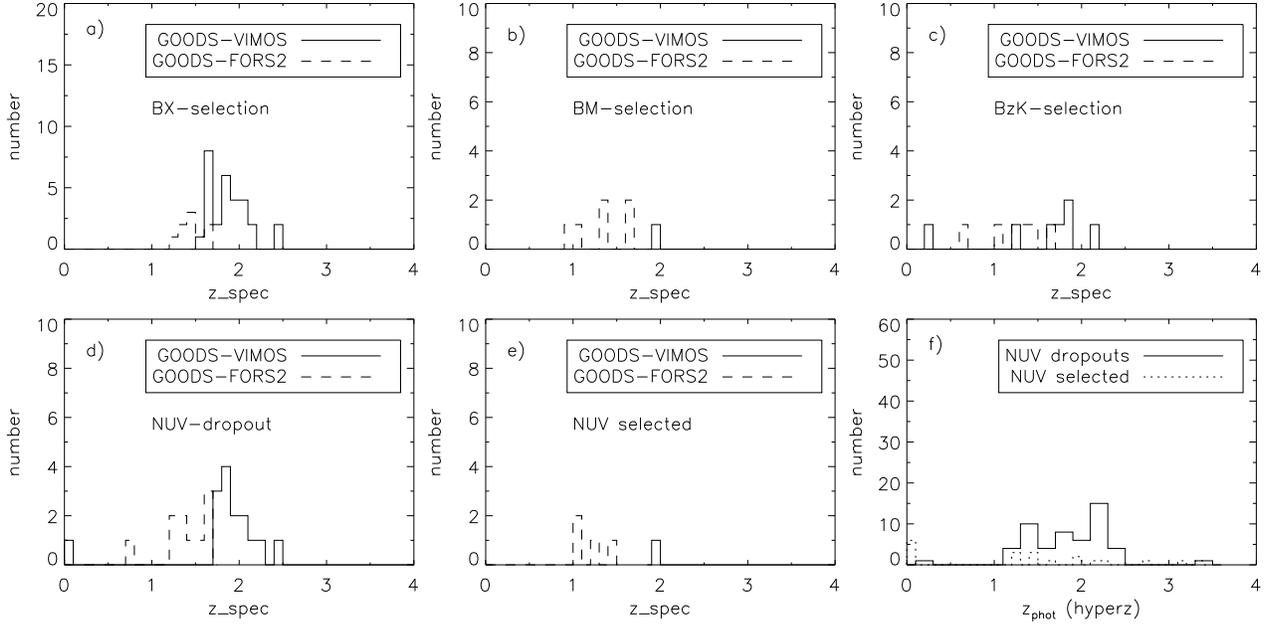}
\caption{\label{spechisto1}Distribution of spectroscopic redshifts,
  panels a) to e), and 
  photometric redshifts, panel f) of BM/BX, BzK, and NUV dropout and 
  NUV-selected star-forming galaxies in the GOODS-S
  field. Spectroscopic redshifts 
  are determined from VLT observations with either the VIMOS (solid
  line) or FORS (dashed line) spectrographs.}
\end{center}
\end{figure*}

These results show that a true Lyman break selection targets a
different population of galaxies, perhaps more evolved and/or reddened
than the BM/BX-selected galaxy population and more similar to the BzK
selection. Our dropout selection targets mainly bright
star forming galaxies due to limitations in the sensitivity, especially in the
K-band. However, compared to the $BzK$ selection, the dropout technique
reduces the contamination due to interlopers and results in a cleaner, more
complete sample of $1.5\le z\le 2.5$ LBGs (Figure~\ref{spechisto1}).  

\subsection{Masses, SFRs, and Extinction}
\begin{figure*}[ht]
\begin{center}
\includegraphics[width=180mm]{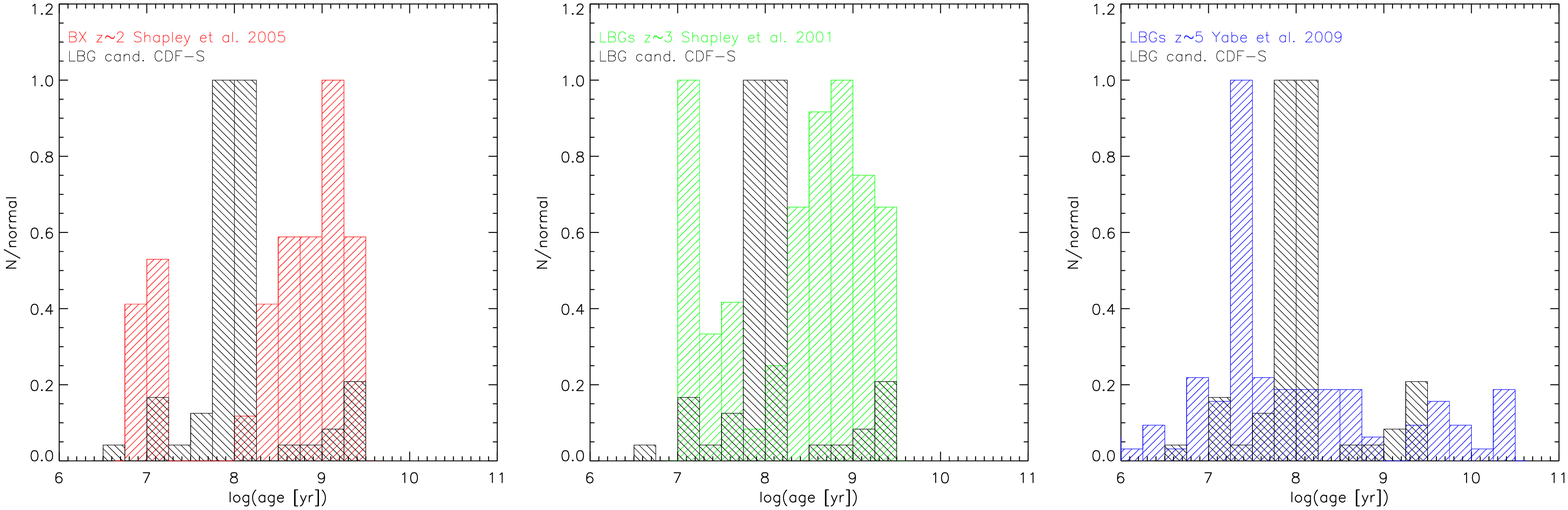}
\caption{\label{age} Comparison of the peak normalized distribution of ages
  between our dropout-selected LBGs (black shaded), BX-selected galaxies of
  \citep[][red shaded bars]{2005ApJ...626..698S} in the left panel, LBGs at
  $z\sim 3$ \citep[][green shaded bars]{2001ApJ...562...95S} in the middle
  panel and LBGs $z\sim 5$ \citep[][blue shaded bars]{2009ApJ...693..507Y} in
  the right panel. The samples used here consist of: this work: 61 NUV
  dropout+NUV selected, Shapley $z\sim 2$: 72; Shapley $z\sim 3$: 73; Yabe
  $z\sim 5$: 105 objects.}
\end{center}
\end{figure*}

\begin{figure*}[ht]
\begin{center}
\includegraphics[width=84mm]{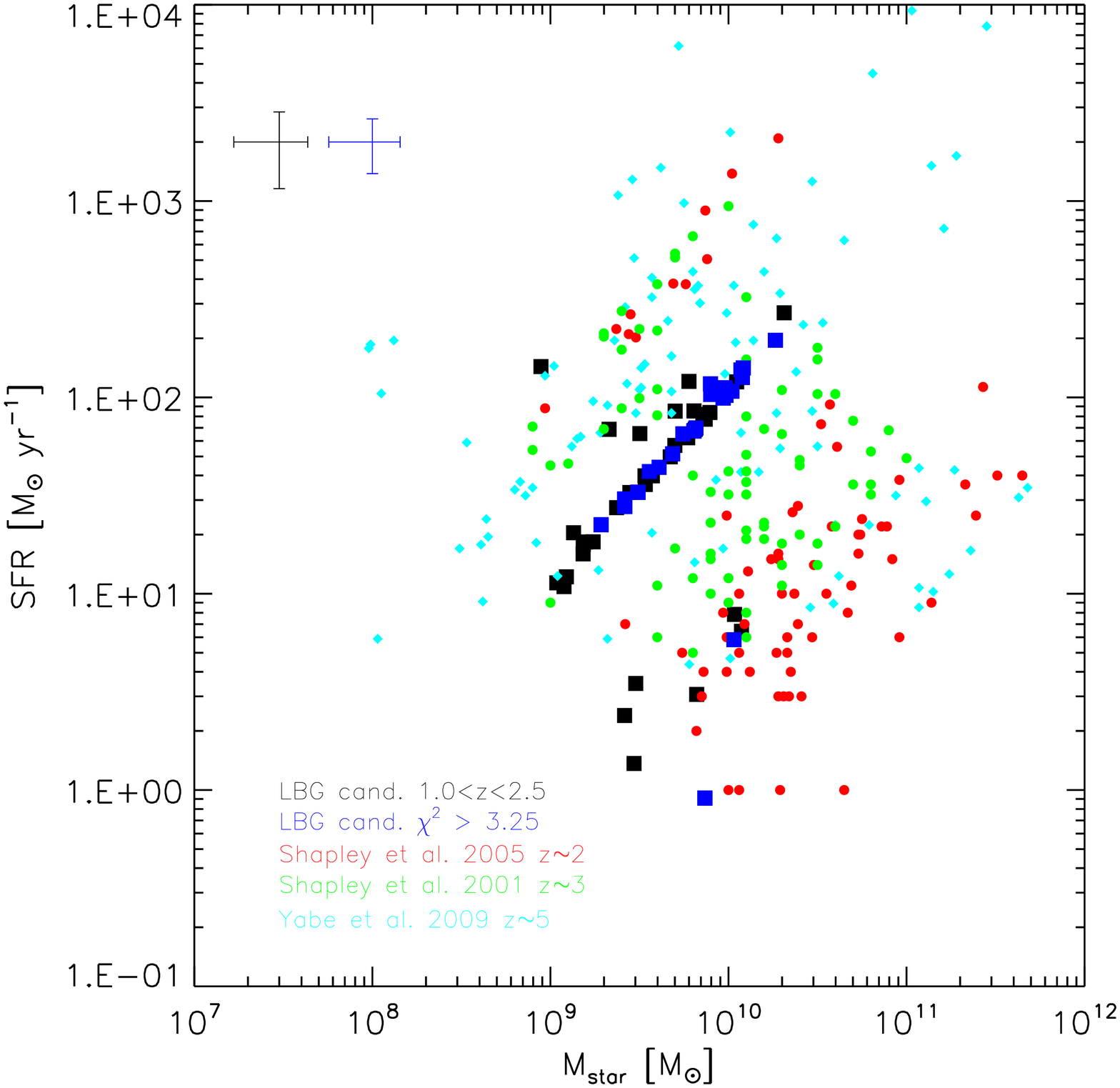}
\includegraphics[width=84mm]{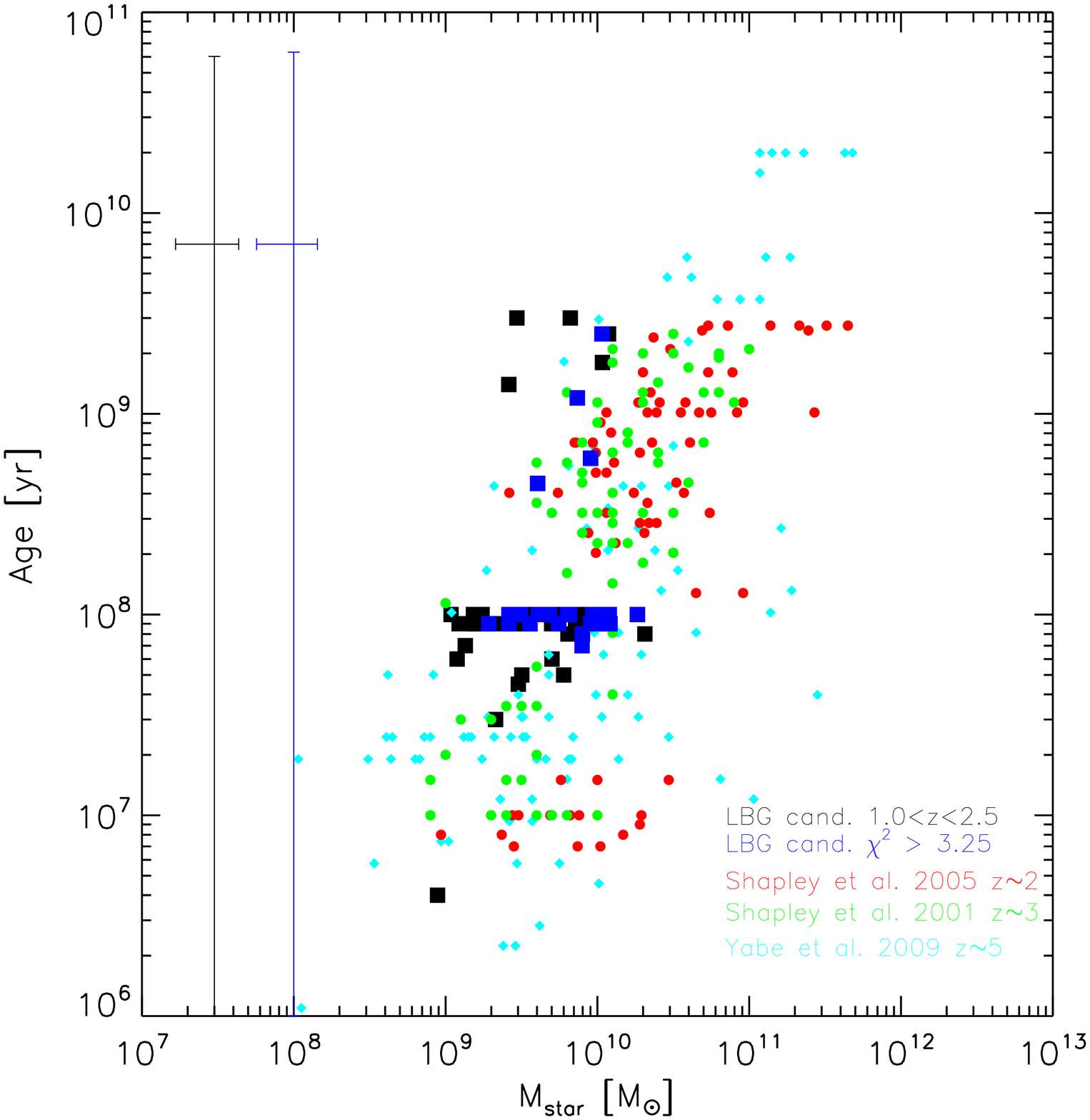}

\caption{\label{mass_sfr} Comparison of masses and SFRs between our 
  dropout-selected $z\sim 2$ LBGs {\it (black and blue squares)}, 
  the $z\sim 2$ BX-selected star-forming galaxies {\it (red filled circles)},
  the $z\sim 3$ LBGs of \citeauthor{2005ApJ...626..698S} {\it (green filled
    circles)}, and $z\sim 5$ LBGs of \citet[][]{2009ApJ...693..507Y} {\it
    (cyan diamonds)}. For high star formation rate systems,
  the Lyman break-selected galaxies have higher masses than the BM/BX
  sample for a given SFR. The error bars in the upper left corner
  indicate typical errors for stellar masses, SFRs, and ages.}
\end{center}
\end{figure*}

Fitting the PEGASE synthesis evolution models, we find that
our UV-bright LBG sample is best represented by ages ranging between
$0.004-3$~Gyr (Figure~\ref{age}), stellar 
masses between $9\times 10^{8}$-$2\times 10^{10}~M_{\odot}$,
and SFRs up to 270~$M_{\odot}~$~yr$^{-1}$
(Figure~\ref{mass_sfr}) with a median  ${\rm SFR = 50\pm 53
  M_{\odot}}$~yr$^{-1}$. The sample is relatively 
strongly reddened, with a median $A_V = 1.1\pm 0.5$. The age
distribution peaks around $0.1$~Gyr (Figure~\ref{age}), with a
possible second smaller peak around $2-3$~Gyr. We see a lack of
the very young galaxies $\lsim 10$~Myr found 
by \citet[][]{2005ApJ...626..698S}, indicating that our color selection
compared to the BX selected star-forming galaxies at $z \sim 2$ of
\citeauthor[][]{2005ApJ...626..698S} is more sensitive to evolved galaxies with
moderate star formation rates.   
However, when comparing the stellar mass and SFR distribution, we find that
the dropout technique is more sensitive to moderate masses with a peak around
$\sim 10^{10}~M_{\odot}$ and SFRs higher than those at $z\sim 3$
\citep[e.g.][]{2001ApJ...562...95S}, peaking at relatively high SFRs of
about $100~M_{\odot}~$~yr$^{-1}$. 
Similar to \citet[][]{2007MNRAS.377.1024V}, we find
that our $z\sim 2$ LBGs are more massive, older, and have similar SFRs
compared to high redshift LBGs, for example the $z\sim 5$ LBGs of
\citet[][]{2009ApJ...693..507Y}.  

In general, the properties of our dropout-selected
LBGs based on the SED fits are more comparable to LBGs at $z\sim 3$
\citep[e.g.][]{2001ApJ...562...95S} than are BM/BX and BzK-selected
galaxies. Our results demonstrate the consistent
selection of the dropout technique over a large redshift range, which 
allows for the compilation of comparable samples to study the evolution
of galaxies over cosmic time scales.


\label{illbg}

\subsection{IR Luminous LBGs}
\begin{figure*}
\includegraphics[width=84mm]{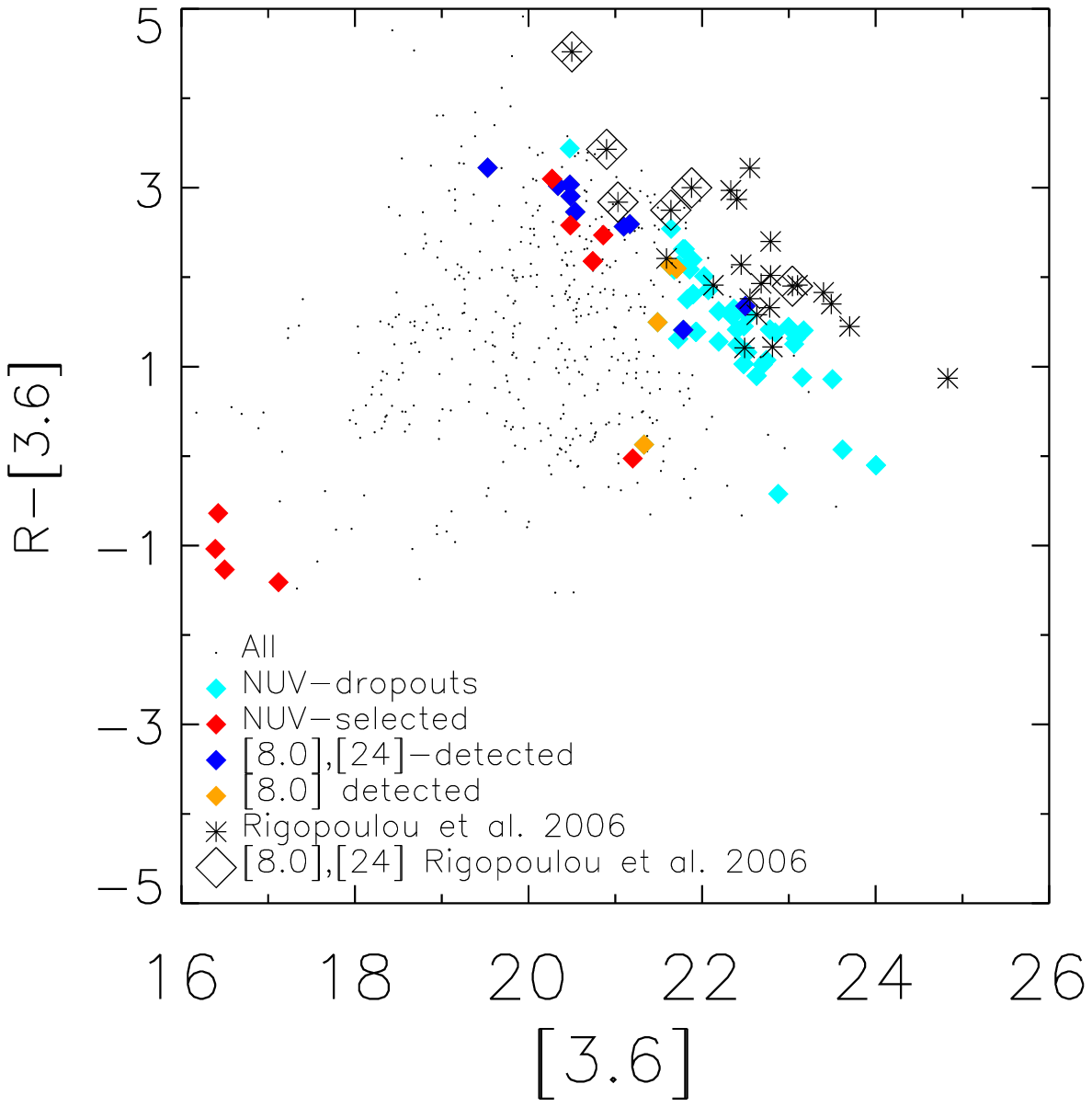}
\includegraphics[width=84mm]{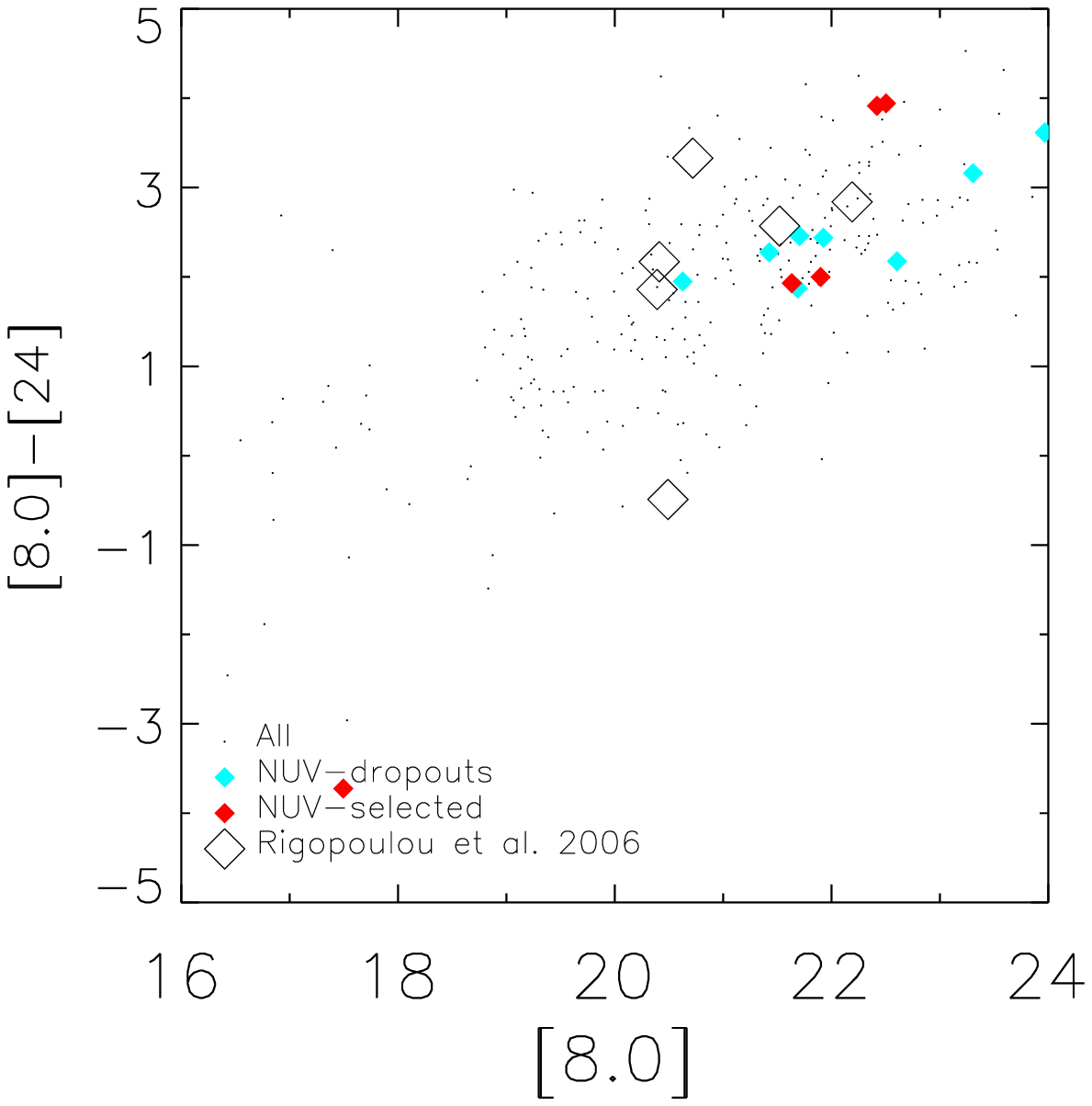}
\caption{\label{spitzer}$[3.6]$ vs $R-[3.6]$ (left panel) color-magnitude
  diagram using Spitzer measurements for our LBG candidate sample. 
  {\it Black dots:} All galaxies in our sample with either IRAC $3.6~\mu$m
  (left panel) or IRAC $8.0~\mu$m and MIPS $24~\mu$m (right panel)
  detections. We compare our measurements to results for $z\sim 3$ ILLBGs
  reported by \citet[][open diamonds]{2006ApJ...648...81R}. We also make a
  similar comparison in the $[8.0]-[24]$ vs $[8.0]$ color-magnitude diagram
  (right panel). {\it Orange diamonds, left panel:}
  LBGs from our sample which are also detected in the Spitzer 8$\mu$m- but not
  in the 24$\mu$m-band.  {\it Open diamonds without open stars, right
    panel:} LBGs from the $z\sim 3$ LBG sample of
  \citet[][]{2006ApJ...648...81R} detected in 8$\mu$m but not 24$\mu$m.}
\end{figure*}

A small fraction of our LBG candidate sample has colors in the $R-[3.6]$
vs. $[3.6]$ color-magnitude diagram (Figure~\ref{spitzer}) comparable to
infrared luminous LBGs (ILLBGs) and sub-millimeter galaxies at $z\sim 3$ 
\citep[e.g.][]{2006ApJ...648...81R}. The location in the color-magnitude
diagrams indicates that our LBG sample consists of about 20\% ILLBGs and/or
galaxies consistent with sub-millimeter galaxies. The estimated median stellar
mass of the IR luminous subsample is $7.9\pm
  4.1\times10^{9}\,M_{\odot}$, and is slightly higher than the median
stellar mass of the total LBG sample ($ 5\pm
    4.1\times10^{9}\,M_{\odot}$). With a median SFR
  of $84\,\pm\,62\,M_{\odot}$~yr$^{-1}$, the IR luminous
  subsample consists mainly of LBGs with high SFRs
  ($\gsim$100~M$_\odot$yr$^{-1}$), while the distribution for the
  general LBG sample show a tail of LBGs with lower SFRs
  ($\lsim$30~M$_\odot$yr$^{-1}$).   
The higher median extinction of $A_V = 1.4\pm 0.8$ for the ILLBGs
($A_V = 1.1\pm 0.5$ for the complete sample) indicates that light can
be subjected to opacities well beyond $\tau = 1$, even in the optical. The
implication is that there is still some part of the star
formation that could be hidden by dust,
resulting in an underestimate of the SFRs. The high dust
  reddening of the ILLBGs is also supported by the fact that all
IR luminous candidates are found with
relatively red colors in the color-color diagrams, not falling in the overlap
region between BM/BX and dropouts \citep[$\left< U-B \right> = 0.72\pm 0.21$ and
$\left< B-R \right> = 0.53\pm 0.27$ consistent with recent results for
a small far-IR detected sample by][]{2011ApJ...734L..12B}. 
Although our results indicate substantial reddening due to larger
amounts of dust, we have to be careful when drawing conclusions
concerning extinction from UV and optical data alone. The
age-extinction degeneracy and the lack of FIR data, where the thermal
emission of dust defines the spectral properties, makes any
interpretation difficult at best. Using the results from the best
fitting SEDs of our sample, we determine a median error for the
extinction of $\Delta
A_V\sim 0.5$. In all cases we find no acceptable fits
with A$_V >$2.3~mag, with the largest extinction corrections being
found for the ILLBGs.

We also calculated the total IR luminosity following the approach of
\citet[][]{2009ApJ...692..556R}, estimating $L_{TIR}$ from the $24~\mu$m flux
using a linear approximation.
The luminosities are given in $L_\odot$, employing apparent solar magnitudes
using \citet[][]{2008AJ....135.2245R} and
\citet[][]{2010AJ....140.1919E}. 
We thus estimated total IR luminosities for our subsample of
ILLBGs of $6\times 10^{10}-1\times 10^{12} L_{\odot}$, with a median
total IR luminosity of $L_{TIR} = 4\times 10^{11} L_{\odot}$. Using the
conversion reported by \cite[][]{2010A&A...518L..24N} this translates into 
a factor of two lower median SFR of about
$40\,M_{\odot}$~yr$^{-1}$ for the ILLBG sample than derived from the
SED model fits.   

Compared to LBG samples at higher and lower redshifts with similar
mass and UV luminosity ranges, we find that our sample consists
of galaxies which are moderately higher extinction. Studies at $z\sim 5$
and $z\sim 3$ estimate median extinctions of $A_V \approx 0.3 - 0.9$
\citep[e.g.][]{2007MNRAS.377.1024V,2009ApJ...693..507Y} and $A_V =
0.5$ \citep[e.g.][]{2001ApJ...562...95S} respectively.  At $z\sim 3$
the ILLBG sample of \citet[][]{2006ApJ...648...81R} shows a median
extinction of $A_V \sim$1.0. Although slightly higher than the general
samples of LBGs at this redshift, their reddening is lower than the
median extinction of $A_V = 1.4$ found in this study. Extinction values
($A_V = 0.8 - 1.8$) similar to ours have
been estimated for a 8$\mu$m and 24$\mu$m detected sample of LBGs by
\citet[][]{2007ApJ...669..749S}. However, their sample consists mainly
of ULIRG-like, massive LBGs with stellar masses of up $10^{12}$M$_\odot$
and IR luminosities of $L_{IR}\gsim 10^{12}L_\odot$.  
Thus our selection, which in general selects $z\sim 2$ LBGs, nevertheless
favors selecting galaxies with higher extinction compared to $z\geq 3$
LBGs.  This is because we
require them to be detected in all bands, including in the IR.

At lower redshifts, studies revealed LBGs with extinction
values similar to the ones found at $z\gsim 3$. At $z\lsim 1$
\citet[][]{2011A&A...526A..10N} estimated extinctions of $A_V=0.5-2$,
with a median of 0.8 for a sample of LBGs in the GOODS-S field
\citep[][]{2007MNRAS.380..986B}. Similarly, at $z\sim 0.1 - 0.2$, Lyman
Break Analogs (LBAs) are found to have a median extinction of $A_V = 0.8$
\citep[][]{2011ApJ...726L...7O}.  While overall trends in the extinction
with redshift are difficult to determine, it appears that our study
stands out in selecting galaxies with moderately higher extinction on
average than other samples at lower and higher redshifts.

%
%
%
%
%
%

Compared with other published examples of IR luminous LBGs, we find that our
sample consists of galaxies with moderate masses having a median mass
of $\sim$8$\times$10$^9$M$_{\odot}$. This is comparable to the $z\sim
1$ LBGs of \citet[][]{2011A&A...526A..10N}, which have masses between
$1.3\times 10^{9}$M$_\odot$ and $3.6\times 10^{10}$M$_\odot$. At $z\sim 3$,
ILLBGs appear to be more massive. \citet[][]{2006ApJ...648...81R},
for example, found IR bright LBGs with masses between $4\times
10^{10}$M$_\odot$ and $4\times 10^{11}$M$_\odot$. Even higher masses
of $10^{12}$M$_\odot$ have been reported for 24$\mu$m-detected 
LBGs by \citet[][]{2007ApJ...669..749S}. With IR luminosities
of $L_{IR}\gsim 10^{12}L_\odot$, these galaxes are massive, dusty ULIRGs.
The very high mass LBGs are relatively rare and might have been missed by
us due to our relatively small search volume. Also, the lack of very 
massive LBGs
in our sample might have been due to limiting our selection to UV-bright
objects, requiring detections from U to IRAC 4.5$\mu$m (observed frame). Both
of these criteria (UV and IR luminous) 
favor the selection of objects with patchy extinction.
Such galaxies possess strong, obscured starbursts.  However, the extinction
is patchy such that light from a UV-emitting stellar population, 
or less obscured
star formation. can be observed.  This can be seen in 
Fig.~\ref{morph_ex}, where some of
the galaxies have a complex morphology.  The ILLBGs represent the more
massive, dusty, heavily star-forming tail of the LBG population in our
study, but are less massive than their high SFR, dusty counterparts 
at higher redshifts.

These galaxies also might resemble an interesting part of the
high redshift sources contributing to cosmic IR background (CIB)
light. At $z\sim 2$ about 30-40\% of 
the CIB results from LIRGs \citep[][]{2010A&A...518L..30B}, which are
comparable to our ILLBGs. In addition, 
far-infrared observations show that at $\lambda
\geq 350\mu$m more than half of the contribution to the CIB  comes from
sources at redshifts $z > 1.2$ \citep[][]{2009Natur.458..737D}. Thus,
a significant fraction of the SFR at this epoch might be provided by dusty
IR-bright galaxies. 

\subsection{Morphology}

\begin{figure*}[ht]
\begin{center}
\includegraphics[width=32mm]{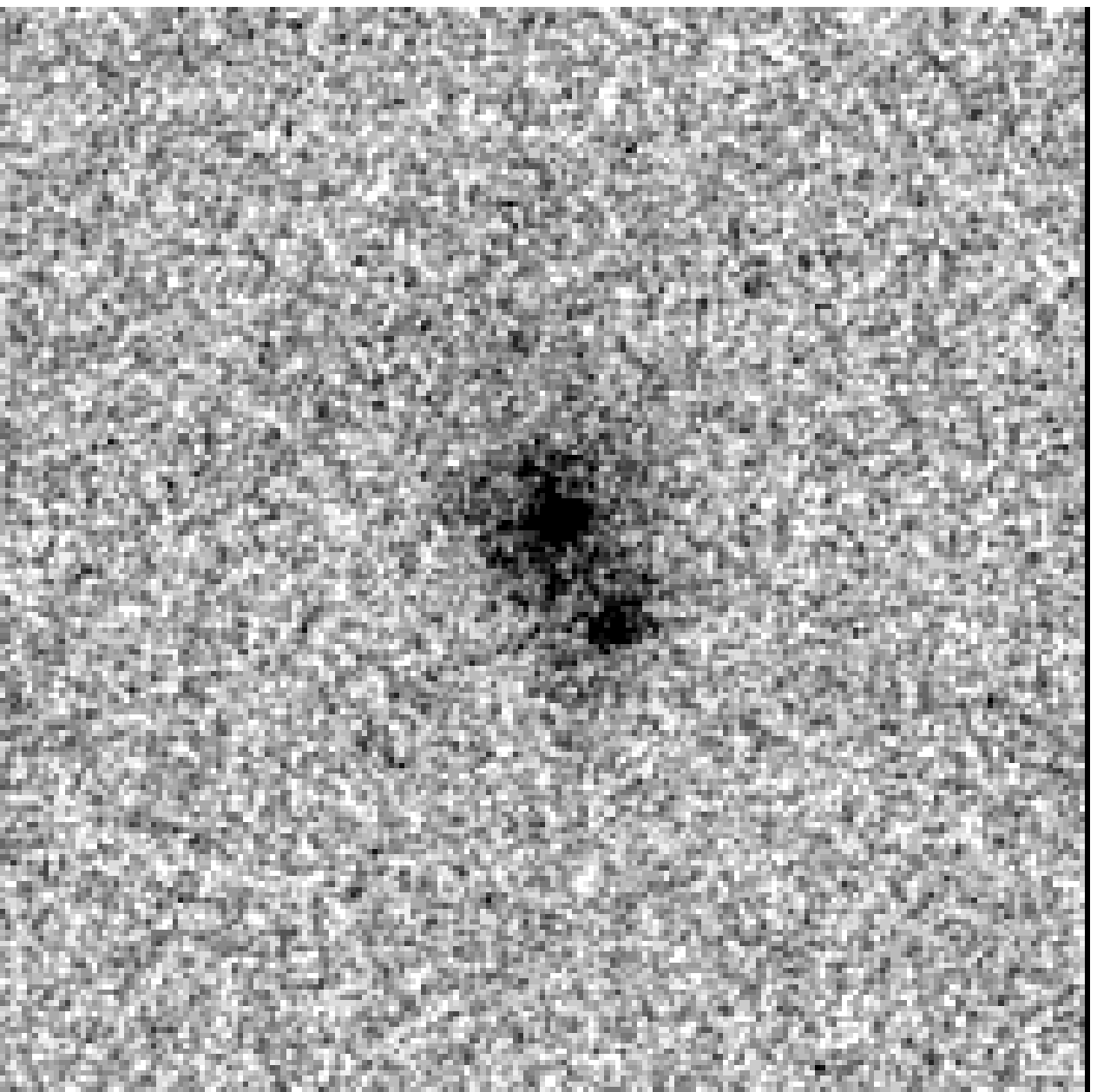}
\hspace{0.1cm}
\includegraphics[width=32mm]{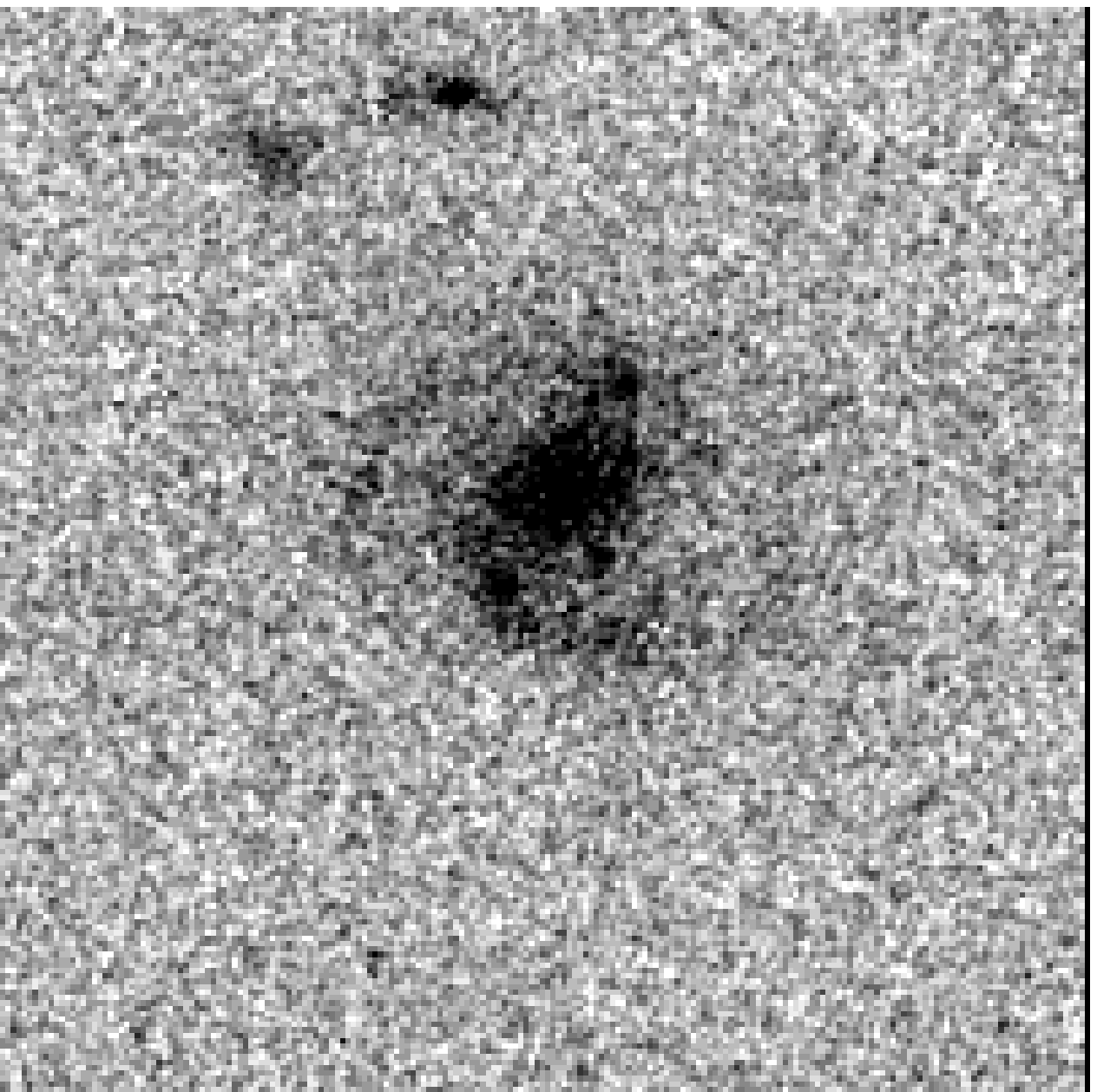}
\hspace{0.1cm}
\includegraphics[width=32mm]{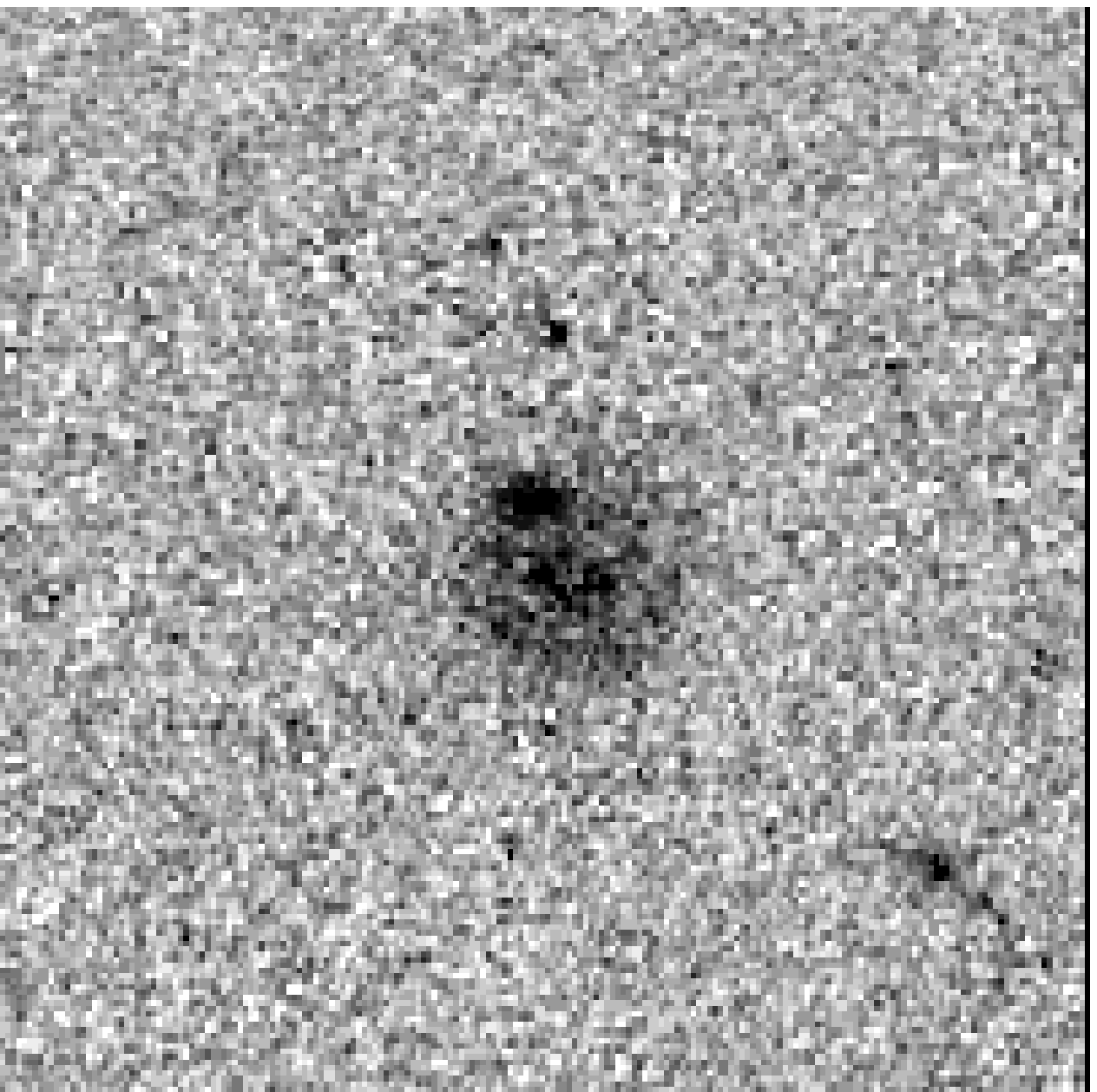}
\hspace{0.1cm}
\includegraphics[width=32mm]{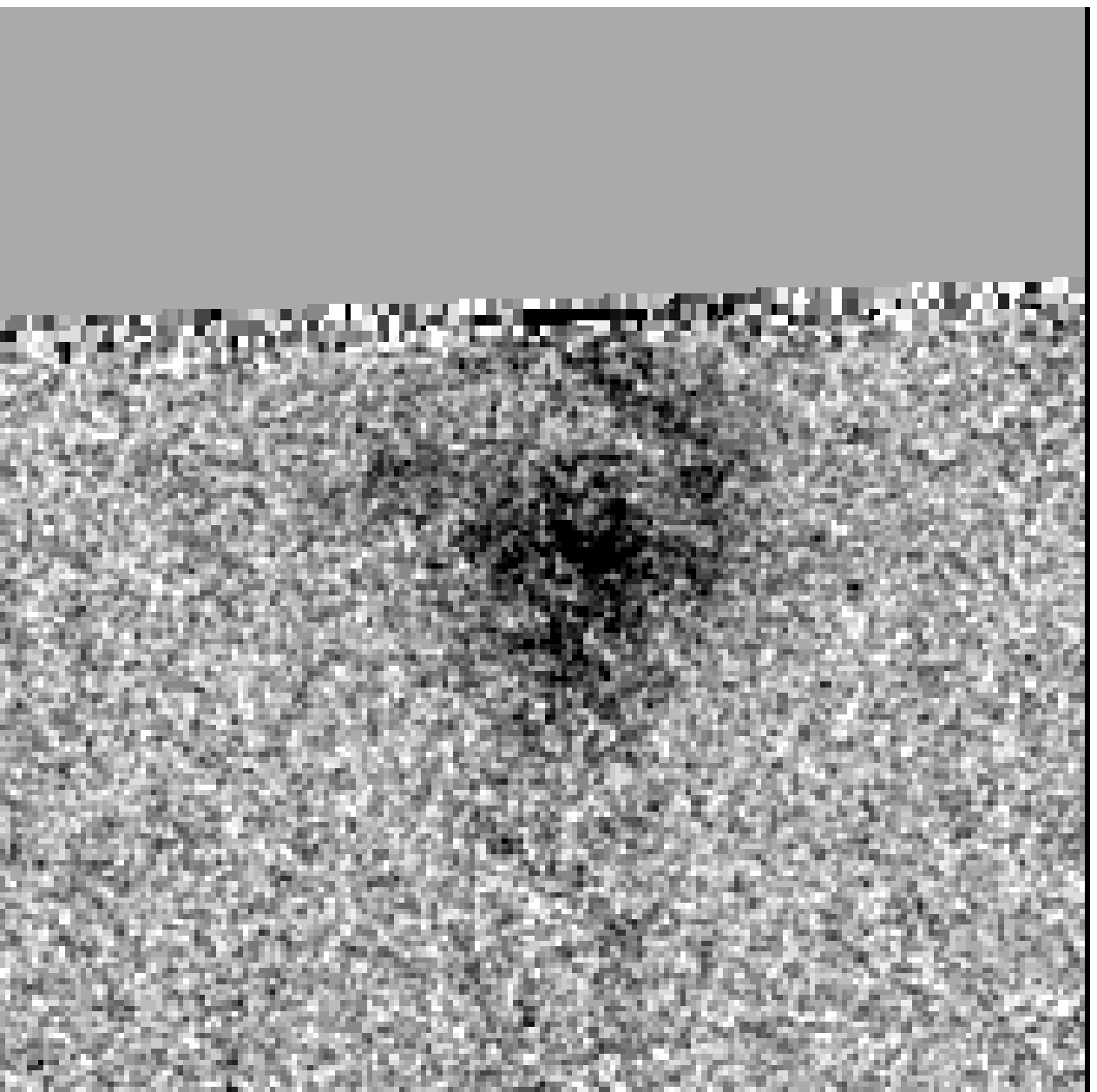}\\
\vspace{0.2cm}
\includegraphics[width=32mm]{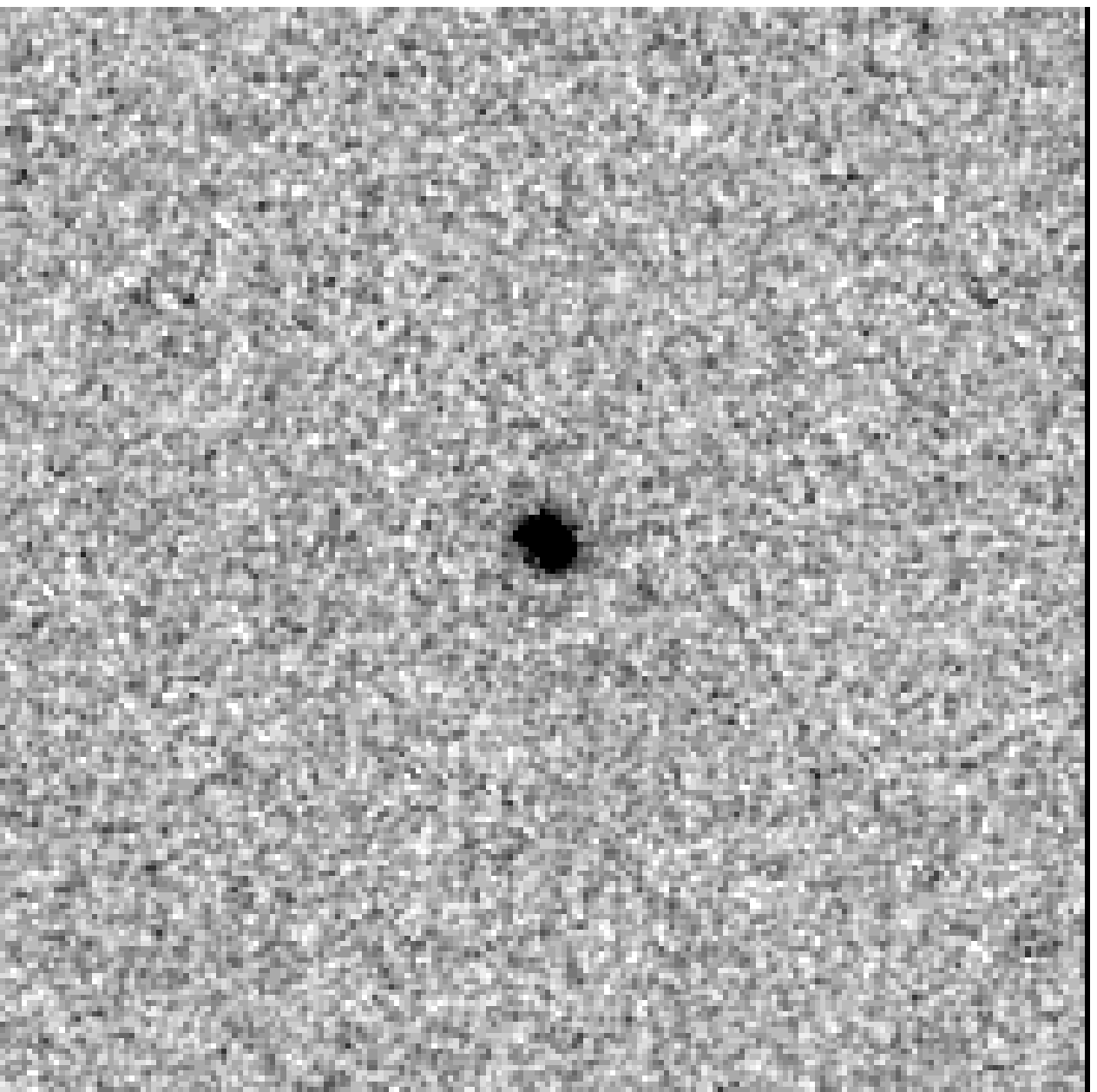}
\hspace{0.1cm}
\includegraphics[width=32mm]{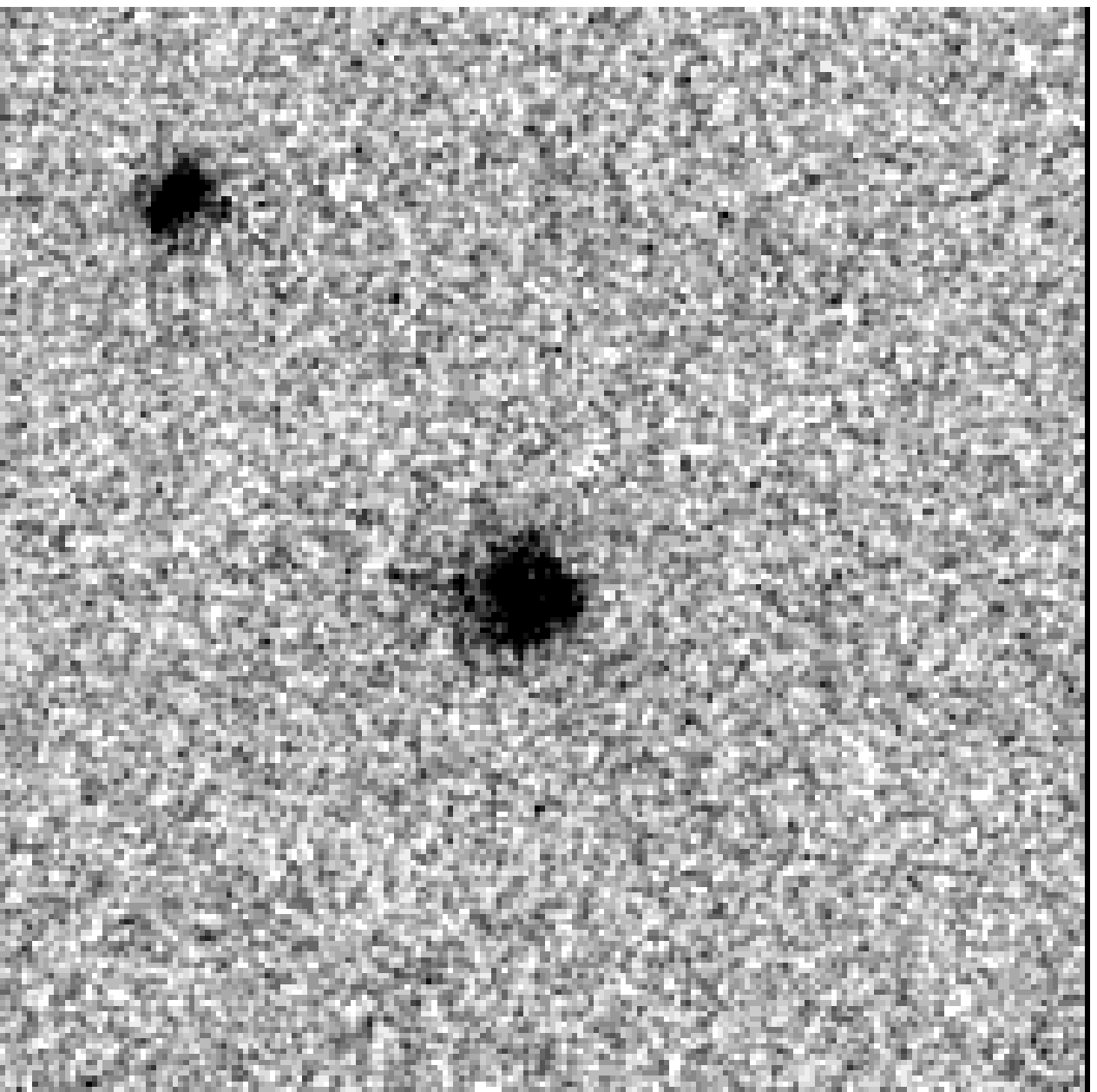}
\hspace{0.1cm}
\includegraphics[width=32mm]{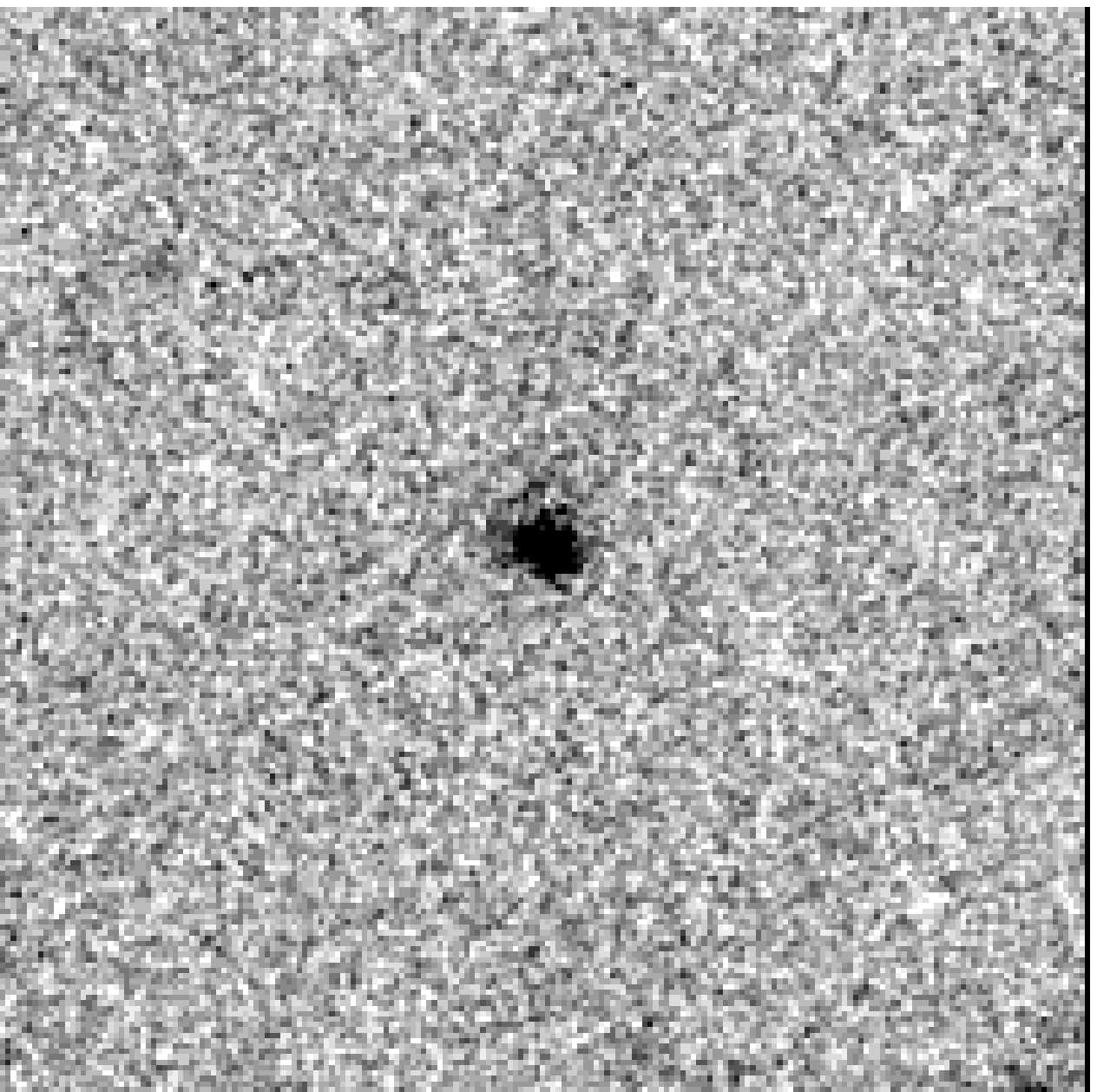}
\hspace{0.1cm}
\includegraphics[width=32mm]{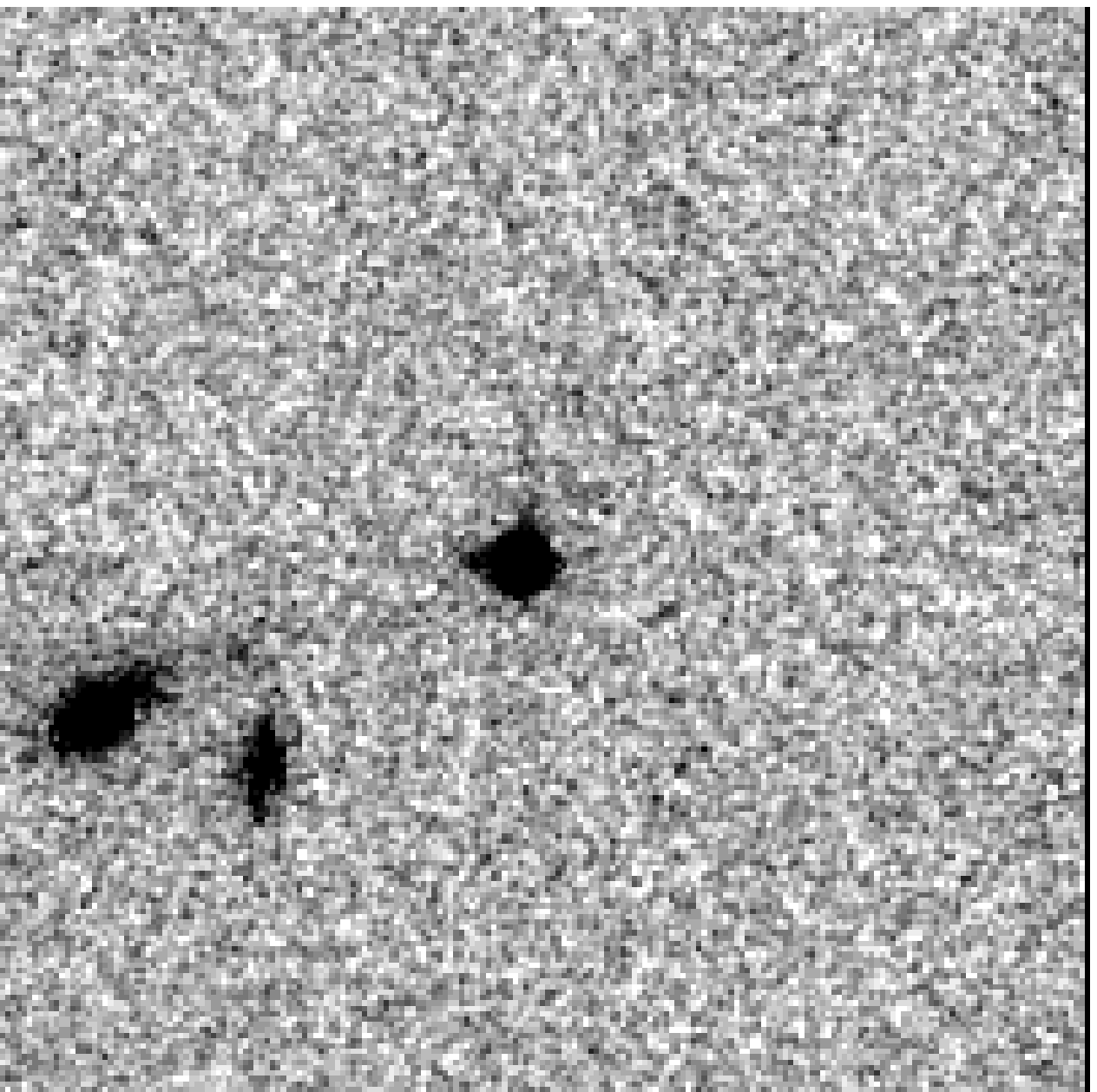}
\caption{\label{morph_ex} Example images showing the morphological
  classification applied. The image size is
  $3^{\prime\prime}\times3^{\prime\prime}$. From left to right, and
referring to galaxies by catalog number: {\it disk-like
    (top row):} \#11800 (dropout-selected LBG), \#7312 (dropout-selected
  ILLBG), \#6039 (NUV-selected LBG), \#7479 (NUV-selected ILLBG);
  {\it compact (bottom row):} \#6633 (dropout-selected LBG), \#12478 
  (dropout-selected LBG), \#7312 (dropout-selected ILLBG), 
  \#10869 (NUV-selected LBG).}
\end{center}
\end{figure*}

From an inspection of the high resolution HST-ACS images by eye, we conclude
that the majority of the LBG sample consists of galaxies showing indications of
disk-like structure, while the rest can be considered compact
objects (examples shown in Figure~\ref{morph_ex}). Also, the majority of the
sample has at least one close 
($< 3^{\prime\prime}$) neighbor or shows disturbed disk structures, such that
they may be considered merger candidates. For the ILLBG subsample,
we find that about 80\% show indication of a disk structure and 20\% are
found to be compact objects. About 50\% of the ILLBGs can be considered merger
candidates. Although these results have to be taken carefully, since we did
not apply detailed morphological analysis and a judgment by eye is very
subjective, they seem to be in agreement with
results found by 
\citet[][]{2006A&A...450...69B} for the $z\sim 1$ LBGs which consist of about
75\% disk-dominated galaxies and $z\sim 1$ LIRGs
\citep[e.g.][]{2005ApJ...632L..65M,2005ApJ...625...23B,2007ApJ...662..322H}. A
more detailed analysis of the morphology will be presented in a later paper
(Haberzettl et al. in prep.).

\section{Summary and Conclusion}

We present results of a search for Lyman break galaxies at 
$1.5\leq z \leq 2.5$, an epoch once called the redshift desert. 
Over the last few years, studies employing the the BM/BX-
\citep[][]{2004ApJ...604..534S,2004ApJ...607..226A} and BzK-methods
\citep[][]{2004ApJ...617..746D,2004ApJ...600L.127D} used observed-frame optical
color-selection criteria to find and study
star-forming galaxies in this redshift range -- which excluded the Lyman break.
Such methods identify galaxies which, though star-forming, are not completely
consistent with LBG samples at higher redshift ($z\gsim 3$). The difference
is that the dropout method allows for a wider range in color redward of
the Lyman limit, which in practice means including redder, more evolved
and potentially more massive galaxies.

Our NUV-dropout plus optical color selection technique targeted 
star-forming galaxies at $1.5\leq z \leq 2.5$, and is similar to the one used by
\citet[][]{2009ApJ...697.1410L}. Using the deep publicly available data set of
the GOODS-S (CDF-S) field covering the UV to mid-IR, we identified a
sample of 73 LBG candidates. 
Our selection method resulted in a success rate of $\sim 80$\%
($\sim$95\% for $1.0\leq z \leq 2.5$). 

The NUV dropout 
technique allows for the detection of more evolved and/or dust-reddened 
galaxies not detected with the BM/BX-method, while also decreasing the
number of interlopers compared to the BzK-method. Although the BM/BX and BzK
methods can select much fainter objects than we can here, due to the
limitation of GALEX in sensitivity, they exclude the more evolved and/or
dust-reddened galaxies completely. The fact that 40\% of the dropout-selected
LBGs are not detected by the BM/BX and/or BzK methods provides for additional
UV/optical-selected star formation at $z\sim 2$, especially if hidden by large
amounts of dust.  

We find that our $z\sim 2$ galaxy sample is consistent with LBGs at
$z\gsim 3$, and therefore allows for consistent follow-up studies at redshifts
spanning $z\sim 1.0 - 7$. 

We calculated useful photometric redshifts using \HyperZ\ and included a small set
of spectroscopic redshifts for our LBG candidate sample
(Figure~\ref{redshift}, red filled circles). The normalized rms scatter
for our LBG candidate sample is $\delta z_{rms} = 0.198$, with a
smaller mean offset of $\langle dz_{offset} \rangle = 0.095$.  

Using the combination of spectroscopic (where available) and photometric
redshifts, we fitted synthetic models from a library of PEGASE spectra to our
LBG sample, and determined ages ranging from 4~Myr
to 3~Gyr, masses between $9\times10^8M_\odot$ to about $2\times
10^{10}M_{\odot}$, and SFRs up to $\approx 270~M_\odot $~yr$^{-1}$. 

Our LBG candidates have moderate stellar masses and SFRs consistent
with $z\sim 3$ LBGs \citep[e.g.][]{2001ApJ...562...95S}. Ages and dust
extinction values of our sample are also more consistent with LBGs at
$z\sim 3$, in comparison with the BX-selected galaxies at $z\sim 2$ of
\citep[e.g.][]{2005ApJ...626..698S}, 
although our sample shows a lack of the very young star-forming galaxies found
in the BM/BX and $z\sim 3$ samples (Figure~\ref{age}). 
Our sample represents the 
moderate mass range of the UV luminosity distribution of LBGs at
this redshift. This can  lead to selection biases against objects with
much higher masses. 
We therefore should exercise care when
comparing our results to other surveys.          

We also identified a small fraction of our LBG sample ($\sim$20\%) as
potential ILLBGs and/or sub-millimeter galaxies using Spitzer 8.0 and
24~$\mu$m data. Although there is some overlap with the other $z\sim 2$
LBGs, the $R-[3.6]$ vs. $[3.6]$ plot in Figure~\ref{spitzer} shows that
the ILLBGs (blue and orange diamonds) represent the UV-bright end of the LBG
distribution for a given value of $R-[3.6]$.  Due to their red colors
at optical wavelengths (compared to the BM/BX selected), most of these
galaxies are missed by the BM/BX and BzK methods ($\sim10$\% detected).

The ILLBG SED fits result, on average, in higher total
stellar masses, SFRs, and extinctions than our overall sample. 

 
Compared with higher and lower redshift
LBGs from the literature, we find that our $z\sim 2$ sample has an average
extinction which is higher than LBGs at $\gsim 3$
\citep[e.g][]{2001ApJ...562...95S,2007MNRAS.377.1024V,2009ApJ...693..507Y}.
A similar result holds in comparison with ILLBGs at $z\gsim 3$
\citep[e.g.][]{2006ApJ...648...81R}, with the exception 
of massive, IR luminous, ULIRG-like LBGs which are selected in the
infrared \citep[][]{2007ApJ...669..749S}. We find that the masses and
extinction values are in better agreement with results for $z\sim 1$
LBGs \citep[][]{2011A&A...526A..10N}.

The median total IR luminosity of our subsample of $L_{TIR} = 4\times
10^{11} L_{\odot}$ indicates that these ILLBGs are comprised mainly of 
LIRGs. Only $\sim15$\% (2 out of 15 ILLBGs) are comparable to ULIRGs
with $L_{TIR} \geq 10^{12} L_{\odot}$. From the total IR luminosities,
we estimate a median SFR half that compared to the one derived from
the SED model fits.  The role of ILLBGs in the overall scheme of star
formation at $z>1.2$ and their contribution to the cosmic IR background
merits further investigation, as ILLBGs could contribute significantly.

A simple morphological classification shows that the majority of our
LBG candidates appear disk-like, and many show indications
for  mergers and interactions. Similar results were found for the ILLBGs,
with 80\% having disk-like morphologies. This
is consistent with results published for $z\sim 1$ LBGs
\citep[][]{2006A&A...450...69B} and $z\sim 1$ LIRGs
\citep[e.g.][]{2005ApJ...632L..65M,2005ApJ...625...23B,2007ApJ...662..322H}.

Overall, our study shows that the selection of star-forming galaxies
at $1.5\le z\le 2.5$ using a true Lyman break dropout technique allows
for the selection of a robust sample of LBGs, which can used for comparable
follow-up studies with LBGs at higher redshifts.  A Lyman break
selection like we used here allows redder, more massive LBGs to be
included in comparison to other selection techniques such as the BM/BX
and BzK methods.

\begin{acknowledgements} 
Based on observations made with ESO Telescopes at the La Silla or Paranal
Observatories under program ID 171.A-3045. 
Observations have been carried out using the Very Large Telescope at the ESO
Paranal Observatory under Programme ID 168.A-0485.
This work is based in part on observations made with the Spitzer Space
Telescope, which is operated by the Jet Propulsion Laboratory, California
Institute of Technology under a contract with NASA. 
MDL and NPHN would like to thank the Centre National de la Recherche
Scientifique (CNRS) for their continuing support.
We thank the reviewer for his/her comments, which helped to improve this
manuscript significantly.
\end{acknowledgements}

Facilities: \facility{ESO-VLT, HST(NICMOS)}

\bibliographystyle{aa}
\bibliography{haberzettl_ref}

\begin{thebibliography}{67}
\expandafter\ifx\csname natexlab\endcsname\relax\def\natexlab#1{#1}\fi

\bibitem[{{Adelberger} {et~al.}(2005){Adelberger}, {Steidel}, {Pettini},
  {Shapley}, {Reddy}, \& {Erb}}]{2005ApJ...619..697A}
{Adelberger}, K.~L., {Steidel}, C.~C., {Pettini}, M., {et~al.} 2005, ApJ, 619,
  697

\bibitem[{{Adelberger} {et~al.}(2004){Adelberger}, {Steidel}, {Shapley},
  {Hunt}, {Erb}, {Reddy}, \& {Pettini}}]{2004ApJ...607..226A}
{Adelberger}, K.~L., {Steidel}, C.~C., {Shapley}, A.~E., {et~al.} 2004, ApJ,
  607, 226

\bibitem[{{Bell} {et~al.}(2005){Bell}, {Papovich}, {Wolf}, {Le Floc'h},
  {Caldwell}, {Barden}, {Egami}, {McIntosh}, {Meisenheimer},
  {P{\'e}rez-Gonz{\'a}lez}, {Rieke}, {Rieke}, {Rigby}, \&
  {Rix}}]{2005ApJ...625...23B}
{Bell}, E.~F., {Papovich}, C., {Wolf}, C., {et~al.} 2005, \apj, 625, 23

\bibitem[{{Berta} {et~al.}(2010){Berta}, {Magnelli}, {Lutz}, {Altieri},
  {Aussel}, {Andreani}, {Bauer}, {Bongiovanni}, {Cava}, {Cepa}, {Cimatti},
  {Daddi}, {Dominguez}, {Elbaz}, {Feuchtgruber}, {F{\"o}rster Schreiber},
  {Genzel}, {Gruppioni}, {Katterloher}, {Magdis}, {Maiolino}, {Nordon},
  {P{\'e}rez Garc{\'{\i}}a}, {Poglitsch}, {Popesso}, {Pozzi}, {Riguccini},
  {Rodighiero}, {Saintonge}, {Santini}, {Sanchez-Portal}, {Shao}, {Sturm},
  {Tacconi}, {Valtchanov}, {Wetzstein}, \& {Wieprecht}}]{2010A&A...518L..30B}
{Berta}, S., {Magnelli}, B., {Lutz}, D., {et~al.} 2010, A\&A, 518, L30+

\bibitem[{{Bolzonella} {et~al.}(2000){Bolzonella}, {Miralles}, \&
  {Pell{\'o}}}]{2000A&A...363..476B}
{Bolzonella}, M., {Miralles}, J.-M., \& {Pell{\'o}}, R. 2000, A\&A, 363, 476

\bibitem[{{Bouwens} {et~al.}(2007){Bouwens}, {Illingworth}, {Franx}, \&
  {Ford}}]{2007ApJ...670..928B}
{Bouwens}, R.~J., {Illingworth}, G.~D., {Franx}, M., \& {Ford}, H. 2007, ApJ,
  670, 928

\bibitem[{{Bouwens} {et~al.}(2004){Bouwens}, {Thompson}, {Illingworth},
  {Franx}, {van Dokkum}, {Fan}, {Dickinson}, {Eisenstein}, \&
  {Rieke}}]{2004ApJ...616L..79B}
{Bouwens}, R.~J., {Thompson}, R.~I., {Illingworth}, G.~D., {et~al.} 2004, ApJ
  Lett., 616, L79

\bibitem[{{Bruzual} \& {Charlot}(2003)}]{2003MNRAS.344.1000B}
{Bruzual}, G. \& {Charlot}, S. 2003, MNRAS, 344, 1000

\bibitem[{{Burgarella} {et~al.}(2011){Burgarella}, {Heinis}, {Magdis}, {Auld},
  {Blain}, {Bock}, {Brisbin}, {Buat}, {Chanial}, {Clements}, {Cooray}, {Eales},
  {Franceschini}, {Giovannoli}, {Glenn}, {Gonz{\'a}lez Solares}, {Griffin},
  {Hwang}, {Ilbert}, {Marchetti}, {Mortier}, {Oliver}, {Page}, {Papageorgiou},
  {Pearson}, {P{\'e}rez-Fournon}, {Pohlen}, {Rawlings}, {Raymond},
  {Rigopoulou}, {Rodighiero}, {Roseboom}, {Rowan-Robinson}, {Scott}, {Seymour},
  {Smith}, {Symeonidis}, {Tugwell}, {Vaccari}, {Vieira}, {Viero}, {Vigroux},
  {Wang}, \& {Wright}}]{2011ApJ...734L..12B}
{Burgarella}, D., {Heinis}, S., {Magdis}, G., {et~al.} 2011, \apjl, 734, L12

\bibitem[{{Burgarella} {et~al.}(2007){Burgarella}, {Le Floc'h}, {Takeuchi},
  {Huang}, {Buat}, {Rieke}, \& {Tyler}}]{2007MNRAS.380..986B}
{Burgarella}, D., {Le Floc'h}, E., {Takeuchi}, T.~T., {et~al.} 2007, MNRAS,
  380, 986

\bibitem[{{Burgarella} {et~al.}(2006){Burgarella}, {P{\'e}rez-Gonz{\'a}lez},
  {Tyler}, {Rieke}, {Buat}, {Takeuchi}, {Lauger}, {Arnouts}, {Ilbert},
  {Barlow}, {Bianchi}, {Lee}, {Madore}, {Malina}, {Szalay}, \&
  {Yi}}]{2006A&A...450...69B}
{Burgarella}, D., {P{\'e}rez-Gonz{\'a}lez}, P.~G., {Tyler}, K.~D., {et~al.}
  2006, A\&A, 450, 69

\bibitem[{{Calzetti} {et~al.}(2000){Calzetti}, {Armus}, {Bohlin}, {Kinney},
  {Koornneef}, \& {Storchi-Bergmann}}]{2000ApJ...533..682C}
{Calzetti}, D., {Armus}, L., {Bohlin}, R.~C., {et~al.} 2000, ApJ, 533, 682

\bibitem[{{Daddi} {et~al.}(2004{\natexlab{a}}){Daddi}, {Cimatti}, {Renzini},
  {Fontana}, {Mignoli}, {Pozzetti}, {Tozzi}, \&
  {Zamorani}}]{2004ApJ...617..746D}
{Daddi}, E., {Cimatti}, A., {Renzini}, A., {et~al.} 2004{\natexlab{a}}, ApJ,
  617, 746

\bibitem[{{Daddi} {et~al.}(2004{\natexlab{b}}){Daddi}, {Cimatti}, {Renzini},
  {Vernet}, {Conselice}, {Pozzetti}, {Mignoli}, {Tozzi}, {Broadhurst}, {di
  Serego Alighieri}, {Fontana}, {Nonino}, {Rosati}, \&
  {Zamorani}}]{2004ApJ...600L.127D}
{Daddi}, E., {Cimatti}, A., {Renzini}, A., {et~al.} 2004{\natexlab{b}}, ApJ
  Lett., 600, L127

\bibitem[{{Daddi} {et~al.}(2005){Daddi}, {Dickinson}, {Chary}, {Pope},
  {Morrison}, {Alexander}, {Bauer}, {Brandt}, {Giavalisco}, {Ferguson}, {Lee},
  {Lehmer}, {Papovich}, \& {Renzini}}]{2005ApJ...631L..13D}
{Daddi}, E., {Dickinson}, M., {Chary}, R., {et~al.} 2005, ApJ Lett., 631, L13

\bibitem[{{Devlin} {et~al.}(2009){Devlin}, {Ade}, {Aretxaga}, {Bock}, {Chapin},
  {Griffin}, {Gundersen}, {Halpern}, {Hargrave}, {Hughes}, {Klein}, {Marsden},
  {Martin}, {Mauskopf}, {Moncelsi}, {Netterfield}, {Ngo}, {Olmi}, {Pascale},
  {Patanchon}, {Rex}, {Scott}, {Semisch}, {Thomas}, {Truch}, {Tucker},
  {Tucker}, {Viero}, \& {Wiebe}}]{2009Natur.458..737D}
{Devlin}, M.~J., {Ade}, P.~A.~R., {Aretxaga}, I., {et~al.} 2009, \nat, 458, 737

\bibitem[{{Engelke} {et~al.}(2010){Engelke}, {Price}, \&
  {Kraemer}}]{2010AJ....140.1919E}
{Engelke}, C.~W., {Price}, S.~D., \& {Kraemer}, K.~E. 2010, AJ, 140, 1919

\bibitem[{{Fioc} \& {Rocca-Volmerange}(1997)}]{1997A&A...326..950F}
{Fioc}, M. \& {Rocca-Volmerange}, B. 1997, A\&A, 326, 950

\bibitem[{{F{\"o}rster Schreiber} {et~al.}(2009){F{\"o}rster Schreiber},
  {Genzel}, {Bouch{\'e}}, {Cresci}, {Davies}, {Buschkamp}, {Shapiro},
  {Tacconi}, {Hicks}, {Genel}, {Shapley}, {Erb}, {Steidel}, {Lutz},
  {Eisenhauer}, {Gillessen}, {Sternberg}, {Renzini}, {Cimatti}, {Daddi},
  {Kurk}, {Lilly}, {Kong}, {Lehnert}, {Nesvadba}, {Verma}, {McCracken},
  {Arimoto}, {Mignoli}, \& {Onodera}}]{2009ApJ...706.1364F}
{F{\"o}rster Schreiber}, N.~M., {Genzel}, R., {Bouch{\'e}}, N., {et~al.} 2009,
  ApJ, 706, 1364

\bibitem[{{Giavalisco} {et~al.}(2004{\natexlab{a}}){Giavalisco}, {Dickinson},
  {Ferguson}, {Ravindranath}, {Kretchmer}, {Moustakas}, {Madau}, {Fall},
  {Gardner}, {Livio}, {Papovich}, {Renzini}, {Spinrad}, {Stern}, \&
  {Riess}}]{2004ApJ...600L.103G}
{Giavalisco}, M., {Dickinson}, M., {Ferguson}, H.~C., {et~al.}
  2004{\natexlab{a}}, ApJ Lett., 600, L103

\bibitem[{{Giavalisco} {et~al.}(2004{\natexlab{b}}){Giavalisco}, {Ferguson},
  {Koekemoer}, {Dickinson}, {Alexander}, {Bauer}, {Bergeron}, {Biagetti},
  {Brandt}, {Casertano}, {Cesarsky}, {Chatzichristou}, {Conselice},
  {Cristiani}, {Da Costa}, {Dahlen}, {de Mello}, {Eisenhardt}, {Erben}, {Fall},
  {Fassnacht}, {Fosbury}, {Fruchter}, {Gardner}, {Grogin}, {Hook},
  {Hornschemeier}, {Idzi}, {Jogee}, {Kretchmer}, {Laidler}, {Lee}, {Livio},
  {Lucas}, {Madau}, {Mobasher}, {Moustakas}, {Nonino}, {Padovani}, {Papovich},
  {Park}, {Ravindranath}, {Renzini}, {Richardson}, {Riess}, {Rosati},
  {Schirmer}, {Schreier}, {Somerville}, {Spinrad}, {Stern}, {Stiavelli},
  {Strolger}, {Urry}, {Vandame}, {Williams}, \& {Wolf}}]{2004ApJ...600L..93G}
{Giavalisco}, M., {Ferguson}, H.~C., {Koekemoer}, A.~M., {et~al.}
  2004{\natexlab{b}}, ApJ Lett., 600, L93

\bibitem[{{Groenewegen} {et~al.}(2002){Groenewegen}, {Girardi},
  {Hatziminaoglou}, {Benoist}, {Olsen}, {da Costa}, {Arnouts}, {Madejsky},
  {Mignani}, {Rit{\'e}}, {Sikkema}, {Slijkhuis}, \&
  {Vandame}}]{2002A&A...392..741G}
{Groenewegen}, M.~A.~T., {Girardi}, L., {Hatziminaoglou}, E., {et~al.} 2002,
  A\&A, 392, 741

\bibitem[{{Gwyn} \& {Hartwick}(2005)}]{2005AJ....130.1337G}
{Gwyn}, S.~D.~J. \& {Hartwick}, F.~D.~A. 2005, \aj, 130, 1337

\bibitem[{{Haberzettl} {et~al.}(2009){Haberzettl}, {Williger}, {Lauroesch},
  {Haines}, {Valls-Gabaud}, {Harris}, {Koekemoer}, {Loveday}, {Campusano},
  {Clowes}, {Dav{\'e}}, {Graham}, \& {S{\"o}chting}}]{2009ApJ...702..506H}
{Haberzettl}, L., {Williger}, G.~M., {Lauroesch}, J.~T., {et~al.} 2009, ApJ,
  702, 506

\bibitem[{{Hammer} {et~al.}(2007){Hammer}, {Puech}, {Chemin}, {Flores}, \&
  {Lehnert}}]{2007ApJ...662..322H}
{Hammer}, F., {Puech}, M., {Chemin}, L., {Flores}, H., \& {Lehnert}, M.~D.
  2007, \apj, 662, 322

\bibitem[{{Hathi} {et~al.}(2010){Hathi}, {Ryan}, {Cohen}, {Yan}, {Windhorst},
  {McCarthy}, {O'Connell}, {Koekemoer}, {Rutkowski}, {Balick}, {Bond},
  {Calzetti}, {Disney}, {Dopita}, {Frogel}, {Hall}, {Holtzman}, {Kimble},
  {Paresce}, {Saha}, {Silk}, {Trauger}, {Walker}, {Whitmore}, \&
  {Young}}]{2010ApJ...720.1708H}
{Hathi}, N.~P., {Ryan}, R.~E., {Cohen}, S.~H., {et~al.} 2010, \apj, 720, 1708

\bibitem[{{Hildebrandt} {et~al.}(2005){Hildebrandt}, {Bomans}, {Erben},
  {Schneider}, {Schirmer}, {Czoske}, {Dietrich}, {Schrabback}, {Simon},
  {Dettmar}, {Haberzettl}, {Hetterscheidt}, \& {Cordes}}]{2005A&A...441..905H}
{Hildebrandt}, H., {Bomans}, D.~J., {Erben}, T., {et~al.} 2005, A\&A, 441, 905

\bibitem[{{Hopkins}(2004)}]{2004ApJ...615..209H}
{Hopkins}, A.~M. 2004, ApJ, 615, 209

\bibitem[{{Kashikawa} {et~al.}(2004){Kashikawa}, {Shimasaku}, {Yasuda},
  {Ajiki}, {Akiyama}, {Ando}, {Aoki}, {Doi}, {Fujita}, {Furusawa}, {Hayashino},
  {Iwamuro}, {Iye}, {Karoji}, {Kobayashi}, {Kodaira}, {Kodama}, {Komiyama},
  {Matsuda}, {Miyazaki}, {Mizumoto}, {Morokuma}, {Motohara}, {Murayama},
  {Nagao}, {Nariai}, {Ohta}, {Okamura}, {Ouchi}, {Sasaki}, {Sato}, {Sekiguchi},
  {Shioya}, {Tamura}, {Taniguchi}, {Umemura}, {Yamada}, \&
  {Yoshida}}]{2004PASJ...56.1011K}
{Kashikawa}, N., {Shimasaku}, K., {Yasuda}, N., {et~al.} 2004, PASJ, 56, 1011

\bibitem[{{Kroupa} {et~al.}(1993){Kroupa}, {Tout}, \&
  {Gilmore}}]{1993MNRAS.262..545K}
{Kroupa}, P., {Tout}, C.~A., \& {Gilmore}, G. 1993, \mnras, 262, 545

\bibitem[{{Le Floc'h} {et~al.}(2005){Le Floc'h}, {Papovich}, {Dole}, {Bell},
  {Lagache}, {Rieke}, {Egami}, {P{\'e}rez-Gonz{\'a}lez}, {Alonso-Herrero},
  {Rieke}, {Blaylock}, {Engelbracht}, {Gordon}, {Hines}, {Misselt}, {Morrison},
  \& {Mould}}]{2005ApJ...632..169L}
{Le Floc'h}, E., {Papovich}, C., {Dole}, H., {et~al.} 2005, ApJ, 632, 169

\bibitem[{{Lehnert} \& {Bremer}(2003)}]{2003ApJ...593..630L}
{Lehnert}, M.~D. \& {Bremer}, M. 2003, ApJ, 593, 630

\bibitem[{{Lehnert} {et~al.}(2010){Lehnert}, {Nesvadba}, {Cuby}, {Swinbank},
  {Morris}, {Cl{\'e}ment}, {Evans}, {Bremer}, \& {Basa}}]{2010Natur.467..940L}
{Lehnert}, M.~D., {Nesvadba}, N.~P.~H., {Cuby}, J., {et~al.} 2010, \nat, 467,
  940

\bibitem[{{Ly} {et~al.}(2009){Ly}, {Malkan}, {Treu}, {Woo}, {Currie},
  {Hayashi}, {Kashikawa}, {Motohara}, \& {Yoshida}}]{2009ApJ...697.1410L}
{Ly}, C., {Malkan}, M.~A., {Treu}, T., {et~al.} 2009, ApJ, 697, 1410

\bibitem[{{Madau}(1995)}]{1995ApJ...441...18M}
{Madau}, P. 1995, ApJ, 441, 18

\bibitem[{{Magdis} {et~al.}(2010){Magdis}, {Rigopoulou}, {Huang}, \&
  {Fazio}}]{2010MNRAS.401.1521M}
{Magdis}, G.~E., {Rigopoulou}, D., {Huang}, J., \& {Fazio}, G.~G. 2010, MNRAS,
  401, 1521

\bibitem[{{Martin} {et~al.}(2005){Martin}, {Fanson}, {Schiminovich},
  {Morrissey}, {Friedman}, {Barlow}, {Conrow}, {Grange}, {Jelinsky},
  {Milliard}, {Siegmund}, {Bianchi}, {Byun}, {Donas}, {Forster}, {Heckman},
  {Lee}, {Madore}, {Malina}, {Neff}, {Rich}, {Small}, {Surber}, {Szalay},
  {Welsh}, \& {Wyder}}]{2005ApJ...619L...1M}
{Martin}, D.~C., {Fanson}, J., {Schiminovich}, D., {et~al.} 2005, ApJ Lett.,
  619, L1

\bibitem[{{Melbourne} {et~al.}(2005){Melbourne}, {Koo}, \& {Le
  Floc'h}}]{2005ApJ...632L..65M}
{Melbourne}, J., {Koo}, D.~C., \& {Le Floc'h}, E. 2005, \apjl, 632, L65

\bibitem[{{Nilsson} {et~al.}(2011){Nilsson}, {M{\"o}ller-Nilsson}, {Rosati},
  {Lombardi}, {K{\"u}mmel}, {Kuntschner}, {Walsh}, \&
  {Fosbury}}]{2011A&A...526A..10N}
{Nilsson}, K.~K., {M{\"o}ller-Nilsson}, O., {Rosati}, P., {et~al.} 2011, \aap,
  526, A10+

\bibitem[{{Noll} {et~al.}(2009){Noll}, {Burgarella}, {Giovannoli}, {Buat},
  {Marcillac}, \& {Mu{\~n}oz-Mateos}}]{2009A&A...507.1793N}
{Noll}, S., {Burgarella}, D., {Giovannoli}, E., {et~al.} 2009, \aap, 507, 1793

\bibitem[{{Nordon} {et~al.}(2010){Nordon}, {Lutz}, {Shao}, {Magnelli}, {Berta},
  {Altieri}, {Andreani}, {Aussel}, {Bongiovanni}, {Cava}, {Cepa}, {Cimatti},
  {Daddi}, {Dominguez}, {Elbaz}, {F{\"o}rster Schreiber}, {Genzel}, {Grazian},
  {Magdis}, {Maiolino}, {P{\'e}rez Garc{\'{\i}}a}, {Poglitsch}, {Popesso},
  {Pozzi}, {Riguccini}, {Rodighiero}, {Saintonge}, {Sanchez-Portal}, {Santini},
  {Sturm}, {Tacconi}, {Valtchanov}, {Wetzstein}, \&
  {Wieprecht}}]{2010A&A...518L..24N}
{Nordon}, R., {Lutz}, D., {Shao}, L., {et~al.} 2010, \aap, 518, L24+

\bibitem[{{Overzier} {et~al.}(2011){Overzier}, {Heckman}, {Wang}, {Armus},
  {Buat}, {Howell}, {Meurer}, {Seibert}, {Siana}, {Basu-Zych}, {Charlot},
  {Gon{\c c}alves}, {Martin}, {Neill}, {Rich}, {Salim}, \&
  {Schiminovich}}]{2011ApJ...726L...7O}
{Overzier}, R.~A., {Heckman}, T.~M., {Wang}, J., {et~al.} 2011, \apjl, 726, L7+

\bibitem[{{Popesso} {et~al.}(2009){Popesso}, {Dickinson}, {Nonino}, {Vanzella},
  {Daddi}, {Fosbury}, {Kuntschner}, {Mainieri}, {Cristiani}, {Cesarsky},
  {Giavalisco}, {Renzini}, \& {The Goods Team}}]{2009A&A...494..443P}
{Popesso}, P., {Dickinson}, M., {Nonino}, M., {et~al.} 2009, A\&A, 494, 443

\bibitem[{{Reddy} {et~al.}(2005){Reddy}, {Erb}, {Steidel}, {Shapley},
  {Adelberger}, \& {Pettini}}]{2005ApJ...633..748R}
{Reddy}, N.~A., {Erb}, D.~K., {Steidel}, C.~C., {et~al.} 2005, ApJ, 633, 748

\bibitem[{{Reddy} \& {Steidel}(2004)}]{2004ApJ...603L..13R}
{Reddy}, N.~A. \& {Steidel}, C.~C. 2004, ApJ Lett., 603, L13

\bibitem[{{Reddy} {et~al.}(2008){Reddy}, {Steidel}, {Pettini}, {Adelberger},
  {Shapley}, {Erb}, \& {Dickinson}}]{2008ApJS..175...48R}
{Reddy}, N.~A., {Steidel}, C.~C., {Pettini}, M., {et~al.} 2008, ApJS, 175, 48

\bibitem[{{Rieke} {et~al.}(2009){Rieke}, {Alonso-Herrero}, {Weiner},
  {P{\'e}rez-Gonz{\'a}lez}, {Blaylock}, {Donley}, \&
  {Marcillac}}]{2009ApJ...692..556R}
{Rieke}, G.~H., {Alonso-Herrero}, A., {Weiner}, B.~J., {et~al.} 2009, ApJ, 692,
  556

\bibitem[{{Rieke} {et~al.}(2008){Rieke}, {Blaylock}, {Decin}, {Engelbracht},
  {Ogle}, {Avrett}, {Carpenter}, {Cutri}, {Armus}, {Gordon}, {Gray}, {Hinz},
  {Su}, \& {Willmer}}]{2008AJ....135.2245R}
{Rieke}, G.~H., {Blaylock}, M., {Decin}, L., {et~al.} 2008, AJ, 135, 2245

\bibitem[{{Rigopoulou} {et~al.}(2006){Rigopoulou}, {Huang}, {Papovich},
  {Ashby}, {Barmby}, {Shu}, {Bundy}, {Egami}, {Magdis}, {Smith}, {Willner},
  {Wilson}, \& {Fazio}}]{2006ApJ...648...81R}
{Rigopoulou}, D., {Huang}, J., {Papovich}, C., {et~al.} 2006, ApJ, 648, 81

\bibitem[{{Rix} {et~al.}(2004){Rix}, {Barden}, {Beckwith}, {Bell}, {Borch},
  {Caldwell}, {H{\"a}ussler}, {Jahnke}, {Jogee}, {McIntosh}, {Meisenheimer},
  {Peng}, {Sanchez}, {Somerville}, {Wisotzki}, \& {Wolf}}]{2004ApJS..152..163R}
{Rix}, H.-W., {Barden}, M., {Beckwith}, S.~V.~W., {et~al.} 2004, ApJS, 152, 163

\bibitem[{{Sanders} {et~al.}(2003){Sanders}, {Mazzarella}, {Kim}, {Surace}, \&
  {Soifer}}]{2003AJ....126.1607S}
{Sanders}, D.~B., {Mazzarella}, J.~M., {Kim}, D., {Surace}, J.~A., \& {Soifer},
  B.~T. 2003, AJ, 126, 1607

\bibitem[{{Sawicki} \& {Thompson}(2006)}]{2006ApJ...648..299S}
{Sawicki}, M. \& {Thompson}, D. 2006, ApJ, 648, 299

\bibitem[{{Shapley} {et~al.}(2001){Shapley}, {Steidel}, {Adelberger},
  {Dickinson}, {Giavalisco}, \& {Pettini}}]{2001ApJ...562...95S}
{Shapley}, A.~E., {Steidel}, C.~C., {Adelberger}, K.~L., {et~al.} 2001, ApJ,
  562, 95

\bibitem[{{Shapley} {et~al.}(2005){Shapley}, {Steidel}, {Erb}, {Reddy},
  {Adelberger}, {Pettini}, {Barmby}, \& {Huang}}]{2005ApJ...626..698S}
{Shapley}, A.~E., {Steidel}, C.~C., {Erb}, D.~K., {et~al.} 2005, ApJ, 626, 698

\bibitem[{{Shim} {et~al.}(2007){Shim}, {Im}, {Choi}, {Yan}, \&
  {Storrie-Lombardi}}]{2007ApJ...669..749S}
{Shim}, H., {Im}, M., {Choi}, P., {Yan}, L., \& {Storrie-Lombardi}, L. 2007,
  \apj, 669, 749

\bibitem[{{Stark} {et~al.}(2009){Stark}, {Ellis}, {Bunker}, {Bundy}, {Targett},
  {Benson}, \& {Lacy}}]{2009ApJ...697.1493S}
{Stark}, D.~P., {Ellis}, R.~S., {Bunker}, A., {et~al.} 2009, \apj, 697, 1493

\bibitem[{{Steidel} {et~al.}(1999){Steidel}, {Adelberger}, {Giavalisco},
  {Dickinson}, \& {Pettini}}]{1999ApJ...519....1S}
{Steidel}, C.~C., {Adelberger}, K.~L., {Giavalisco}, M., {Dickinson}, M., \&
  {Pettini}, M. 1999, ApJ, 519, 1

\bibitem[{{Steidel} {et~al.}(2003){Steidel}, {Adelberger}, {Shapley},
  {Pettini}, {Dickinson}, \& {Giavalisco}}]{2003ApJ...592..728S}
{Steidel}, C.~C., {Adelberger}, K.~L., {Shapley}, A.~E., {et~al.} 2003, ApJ,
  592, 728

\bibitem[{{Steidel} {et~al.}(2010){Steidel}, {Erb}, {Shapley}, {Pettini},
  {Reddy}, {Bogosavljevi{\'c}}, {Rudie}, \& {Rakic}}]{2010ApJ...717..289S}
{Steidel}, C.~C., {Erb}, D.~K., {Shapley}, A.~E., {et~al.} 2010, \apj, 717, 289

\bibitem[{{Steidel} \& {Hamilton}(1993)}]{1993AJ....105.2017S}
{Steidel}, C.~C. \& {Hamilton}, D. 1993, AJ, 105, 2017

\bibitem[{{Steidel} {et~al.}(1995){Steidel}, {Pettini}, \&
  {Hamilton}}]{1995AJ....110.2519S}
{Steidel}, C.~C., {Pettini}, M., \& {Hamilton}, D. 1995, AJ, 110, 2519

\bibitem[{{Steidel} {et~al.}(2004){Steidel}, {Shapley}, {Pettini},
  {Adelberger}, {Erb}, {Reddy}, \& {Hunt}}]{2004ApJ...604..534S}
{Steidel}, C.~C., {Shapley}, A.~E., {Pettini}, M., {et~al.} 2004, ApJ, 604, 534

\bibitem[{{Vanzella} {et~al.}(2008){Vanzella}, {Cristiani}, {Dickinson},
  {Giavalisco}, {Kuntschner}, {Haase}, {Nonino}, {Rosati}, {Cesarsky},
  {Ferguson}, {Fosbury}, {Grazian}, {Moustakas}, {Rettura}, {Popesso},
  {Renzini}, {Stern}, \& {The Goods Team}}]{2008A&A...478...83V}
{Vanzella}, E., {Cristiani}, S., {Dickinson}, M., {et~al.} 2008, A\&A, 478, 83

\bibitem[{{Vanzella} {et~al.}(2005){Vanzella}, {Cristiani}, {Dickinson},
  {Kuntschner}, {Moustakas}, {Nonino}, {Rosati}, {Stern}, {Cesarsky}, {Ettori},
  {Ferguson}, {Fosbury}, {Giavalisco}, {Haase}, {Renzini}, {Rettura}, {Serra},
  \& {The Goods Team}}]{2005A&A...434...53V}
{Vanzella}, E., {Cristiani}, S., {Dickinson}, M., {et~al.} 2005, A\&A, 434, 53

\bibitem[{{Vanzella} {et~al.}(2006){Vanzella}, {Cristiani}, {Dickinson},
  {Kuntschner}, {Nonino}, {Rettura}, {Rosati}, {Vernet}, {Cesarsky},
  {Ferguson}, {Fosbury}, {Giavalisco}, {Grazian}, {Haase}, {Moustakas},
  {Popesso}, {Renzini}, {Stern}, \& {The Goods Team}}]{2006A&A...454..423V}
{Vanzella}, E., {Cristiani}, S., {Dickinson}, M., {et~al.} 2006, A\&A, 454, 423

\bibitem[{{Verma} {et~al.}(2007){Verma}, {Lehnert}, {F{\"o}rster Schreiber},
  {Bremer}, \& {Douglas}}]{2007MNRAS.377.1024V}
{Verma}, A., {Lehnert}, M.~D., {F{\"o}rster Schreiber}, N.~M., {Bremer}, M.~N.,
  \& {Douglas}, L. 2007, MNRAS, 377, 1024

\bibitem[{{Yabe} {et~al.}(2009){Yabe}, {Ohta}, {Iwata}, {Sawicki}, {Tamura},
  {Akiyama}, \& {Aoki}}]{2009ApJ...693..507Y}
{Yabe}, K., {Ohta}, K., {Iwata}, I., {et~al.} 2009, ApJ, 693, 507

\end{thebibliography}

\end{document}